\documentclass[10pt,letterpaper]{article}
\usepackage[top=0.85in,left=2.75in,footskip=0.75in]{geometry}

\usepackage{amsmath,amssymb}

\usepackage{changepage}

\usepackage[utf8x]{inputenc}

\usepackage{textcomp,marvosym}

\usepackage{cite}

\usepackage{nameref,hyperref}

\usepackage{microtype}
\DisableLigatures[f]{encoding = *, family = * }

\usepackage[table]{xcolor}

\usepackage{array}

\newcolumntype{+}{!{\vrule width 2pt}}

\newlength\savedwidth

\newcommand\thickhline{\noalign{\global\savedwidth\arrayrulewidth\global\arrayrulewidth 2pt}%
\hline
\noalign{\global\arrayrulewidth\savedwidth}}

\raggedright
\usepackage[aboveskip=1pt,labelfont=bf,labelsep=period,justification=raggedright,singlelinecheck=off]{caption}

\makeatletter
\renewcommand{\@biblabel}[1]{\quad#1.}
\makeatother

\usepackage{lastpage,fancyhdr,graphicx}
\usepackage{epstopdf}
\usepackage{multirow}
\fancyheadoffset[L]{2.25in}
\fancyfootoffset[L]{2.25in}
\begin{document}
\vspace*{0.2in}

\begin{flushleft}
{\Large
\textbf\newline{Associating ridesourcing with road safety outcomes: Insights from Austin, Texas
} 
}
\newline
\\
Eleftheria Kontou\textsuperscript{1,*},
Noreen McDonald\textsuperscript{2}

\bigskip
\textbf{1} Department of Civil and Environmental Engineering, University of Illinois at Urbana-Champaign, Urbana, Illinois, USA
\\
\textbf{2} Department of City and Regional Planning, University of North Carolina at Chapel Hill, Chapel Hill, North Carolina, USA
\\
\bigskip

* Corresponding author. Email: \href{mailto:kontou@illinois.edu}{kontou@illinois.edu}

\end{flushleft}
\section*{Abstract}
Improving road safety and setting targets for reducing traffic-related crashes and deaths are highlighted as part of the United Nations sustainable development goals and worldwide vision zero efforts. The advent of transportation network companies and ridesourcing expands mobility options in cities and may impact road safety outcomes. We analyze the effects of ridesourcing use on road crashes, injuries, fatalities, and driving while intoxicated (DWI) offenses in Travis County, Texas. Our approach leverages real-time ridesourcing volume to explain variation in road safety outcomes. Spatial panel data models with fixed-effects are deployed to examine whether the use of ridesourcing is significantly associated with road crashes and other safety metrics. Our results suggest that for a 10\% increase in ridesourcing trips, we expect a 0.12\% decrease in road crashes, a 0.25\% decrease in road injuries, and a 0.36\% decrease in DWI offenses in Travis County. On the other hand, ridesourcing use is not significantly associated with road fatalities. This study augments existing work because it moves beyond binary indicators of ridesourcing availability and analyzes crash and ridesourcing trips patterns within an urbanized area rather than their metropolitan-level variation. Contributions include developing a data-rich approach for assessing the impacts of ridesourcing use on the transportation system's safety, which may serve as a template for future analyses for other cities. Our findings provide feedback to policymakers by clarifying associations between ridesourcing use and traffic safety and uncover the potential to achieve safer mobility systems with transportation network companies.
\newpage


\section*{Introduction}
Ridesourcing is a term used to describe the operation of transportation network companies such as Uber, Lyft, DiDi Chuxing, and RideAustin, where the rider can hail a vehicle from the convenience of their smartphone using a web or smartphone application. Ridesourcing services are offered in more than a hundred cities in the US and hundreds of cities worldwide. As an example, ridesourcing trips have grown exponentially, with Uber providing ten billion trips globally in 2018~\cite{UberSec}.

Transportation network companies' service can bridge mobility supply gaps, offering a convenient and competitive mode alternative with pooled-service capabilities~\cite{Lavieri2018}, enhancing urban transportation options~\cite{Hall2018}, potentially reducing private vehicle ownership~\cite{Ward2019}, and even providing emergency services by replacing expensive ambulances~\cite{Moskatel2019,Wolfe2020}. However, empirical studies uncover adverse effects of ridesourcing travel, including associations with congestion~\cite{Erhardt2019}, competition with public transportation modes~\cite{Hall2018,Kong2020}, increased net energy use~\cite{Wenzel2019}, and environmental externalities~\cite{Fry2019}. 

Ridesourcing operations could lead to road safety benefits by reducing road crashes, driving while intoxicated (DWI) offenses, and the welfare losses associated with those~\cite{Greenwood2017}. Road safety improvements are a focus of several initiatives around the world, such as the United Nations' sustainable development goals, and in the United States through vision zero efforts~\cite{UNDESA}. Ridesourcing use in a city could be associated with a reduction of DWI offenses in locations where other transport modes are not provided or are less attractive than driving one's vehicle under the influence of alcohol (e.g., use of ridesourcing due to alcohol consumption was highly rated by a survey’s respondents) ~\cite{Goodspeed2019}. On the other hand, ridesourcing use could be positively associated with crashes, based on existing empirical evaluations of the impact of ridesourcing on road safety at the metropolitan level. Existing empirical studies denote that ridesourcing is associated with an overall increase in vehicle miles traveled due to drivers deadheading between offering rides as well as cruising when waiting to be assigned to the next passenger pick-up \cite{Wenzel2019, Kontou2020}. There is also evidence of induced travel contributing to vehicle miles traveled rise~\cite{Lavieri2019} which does not only result in congested city centers but is also related to increased road crash rates~\cite{Barrios2019}.

In this study, we investigate how the use of ridesourcing influences the rate of vehicular crashes, injuries, fatalities, and DWI offenses, and we capture safety-related externalities of ridesourcing use. To our knowledge, this is the first empirical study that uncovers such associations while accounting for the intensity of ridesourcing use by leveraging actual ridesourcing pick-up and drop-off coordinates. Previous evaluations of ridesourcing impacts on road safety outcomes focused on effects across metropolitan regions and used dichotomous indicators of ridesourcing presence or absence as a predictor variable. Instead, we utilize more granular data on ridesourcing trips per census tract for Travis County, Texas. Spatial panel data models are employed to demonstrate associations between ridesourcing trips and road safety outcomes. 

\section*{Literature review}
Existing studies focus on uncovering associations between ridesourcing entry and road safety outcomes, DWI offenses, and alcohol consumption. As shown in Table~\ref{table1}, almost all empirical studies examine the impact of ridesourcing services entry on the relative change in road crash rates (e.g.,~\cite{Brazil2016}). Out of eight economics and epidemiology studies in this field, only one attempts to capture ridesourcing intensity effects. Barrios et al. use Google search count data as a ridesourcing exposure proxy due to the absence of real-world information on this type of trips~\cite{Barrios2019}. Even though safety-related literature highlights significant spatio-temporal associations between traffic and road safety outcomes~\cite{Whitelegg1987,Noland2005}, the studies mentioned above do not capture such effects. Our analysis at a local spatial scale is more appropriate for safety management and policy response than an analysis at a metropolitan scale~\cite{Whitelegg1987}. 

\begin{table}[!ht]
\begin{adjustwidth}{-2.25in}{0in} 
\centering
\caption{
{\bf Summary of methods from existing literature.}}
{\begin{tabular}{|l|p{2.0cm}|p{2.5cm}|p{2.4cm}|p{2.5cm}|p{2.0cm}|p{1.5cm}|}
\hline
\textbf{Reference} & \textbf{Method} & \textbf{Dependent variable} & \textbf{Ridesourcing indicator} & \textbf{Controls} & \textbf{Spatial unit} & \textbf{Time unit}\\ \hline
\thickhline
\cite{Greenwood2017} & DiD, OLS, QMLE & Alcohol-related and total traffic fatalities, alcohol-related deaths on weekends and holidays & UberX/Uber Black launch dates & Age, education, population, median income, poverty rate, law enforcement population & Townships (CA, US) & Quarter (2009-2014) \\ \hline
\cite{Brazil2016} & DiD Poisson, Negative binomial & Total, alcohol-involved, weekend and holiday traffic fatalities & Uber launch date & State laws, state beer tax, unemployment rate & US metro area counties (100 largest) & Month 
(2009-2014) \\ \hline
\cite{MartinBuck} & DiD & Alcohol-related fatal crashes, DUI, and other offenses & Uber and Lyft launch dates & Population, unemployment rate, light-rail transport use  & US Cities:
273 cities of $>100,000$ population & Month (2000-2014) \\ \hline
\cite{Morrison2018} & ARIMA (with transfer functions) & All, alcohol-involved, serious injury and fatality crashes & Uber launch, cease, and resume dates & n.a. & 4 US cities & Week (2013-2016) \\ \hline
\cite{Dills2018} & DiD & Fatal crash rate
(total, alcohol-involved, night-time), offenses rate & UberX launch date & State driving laws, state beer tax, age, race, unemployment rate, population density & US counties & Month (2007-2015) \\ \hline
\cite{Barrios2019} & DiD & Total, drunk, pedestrian-involved, non-drunk crashes and fatalities & Intensity of rideshare use (Uber and Lyft Google search) for conventional and pooled ride service & Population, per capita income, vehicle ownership, public transportation use, VMT, new car registrations, quality of drivers & US cities: population $>10,000$ people & Quarter (2010-2017) \\ \hline
\cite{Huang2019} & DiD, ARIMA & Road traffic deaths & Uber launch date &Age, sex, birth province & South African province & Week (2010-2014) \\ \hline
\cite{Kirk2020} & Negative binomial & Traffic injuries & Uber availability (binary) & Employment rates, fuel price, taxis & Great Britain local government authorities & Monthly (2009-2017) \\ \hline
\end{tabular}}
\begin{flushleft}DiD stands for difference-in-differences, OLS for ordinary least squares, QMLE for Poisson quasi-maximum likelihood estimator, and ARIMA for autoregressive integrated moving average.
\end{flushleft}
\label{table1}
\end{adjustwidth}
\end{table}

Across studies, safety outcomes examined vary between crash, injury, and fatality rates, as well as DWI offense rates. A subset of studies focuses on modeling alcohol-related fatalities and injuries~\cite{Greenwood2017, MartinBuck,Morrison2018}.

Difference-in differences models are the primary approach in the existing literature to capture causal relationships between ridesourcing entry and road safety outcomes but a few researchers also conduct time-series analysis to track trends over time~\cite{Morrison2018,Huang2019}. Among the studies reviewed, the following variables serve as controls: vehicle miles traveled, annual average daily traffic or population are included as traffic measures, and indicators of socio-demographic and economic characteristics, driving-related laws, vehicle ownership, and public transit use. Time units of such analysis vary between weeks, months, or quarters. The smallest spatial unit of the observations used in such analysis is more commonly a city, a county, or a metropolitan statistical area, primarily due to the lack of more granular data of transportation network companies' travel patterns.

Findings are not consistent across studies, even comparing those that follow similar methods and use similar spatio-temporal units of analysis. For example, Brazil and Kirk conclude that transportation network companies entry is not significantly associated with any of the categories of traffic fatalities examined in their work for the US~\cite{Brazil2016}; a similar conclusion is reached by Huang et al.~\cite{Huang2019} in their South African-focused study. Morrison et al. hypothesize that the resumption of ridesourcing operation in specific US cities is associated with a decrease in alcohol-involved road crashes~\cite{Morrison2018}; their hypothesis is partially supported for the cities of Portland, Oregon and San Antonio, Texas. However, the studies above do not account for actual ridesourcing travel patterns, which are much needed to capture local effects by overlaying ridesourcing pick-up and drop-off locations and safety outcomes. Other studies also report findings that are in general agreement with the aforementioned outcomes. As an example, Dills and Mulholland indicate that Uber's launch is associated with a reduction in fatal traffic crashes and DWI offenses after a certain number of months of operation, uncovering lags~\cite{Dills2018}.

On the contrary, Barrios et al. find that road fatalities (including pedestrian and non-vehicle occupants) increase with ridesourcing use and that these trends persist over time~\cite{Barrios2019}. They also demonstrate that the road crash rate increase is significant in cities with greater population levels, higher income quartiles, greater vehicle ownership share and public transportation use, as well as higher carpooling use. Barrios et al. is the only work that captures such exposure, using a proxy for adoption intensity, by adopting Google trends search counts~\cite{Barrios2019}. We find that the existing literature fails to assess how the intensity of ridesourcing services in smaller geographic areas is related to road safety since it does not leverage actual locations of crashes and other road safety outcomes data. 

Ridesourcing operation resembles quite a lot that of a taxi service. Evidence from safety research of the operation of taxis examines the impact of fatigue~\cite{Dalziel1997}, driver behavior, and working conditions~\cite{Wang2019}, as well as taxi driver offenses~\cite{La2013} on taxi crashes. Associations of these factors with taxi crashes and fatalities are examined through Poisson and logistic regression models without taking into account spatial and temporal characteristics. Such models, along with multivariate regressions models (e.g.,~\cite{Gonzalez2019}), seemingly unrelated regressions (e.g.,~\cite{Jones2019}), and geographically weighted regressions (e.g.,~\cite{Huang2018,Wang2016}) are leveraged in a variety of road safety outcomes modeling studies. Note that after systematically searching for evidence, we have not identified studies that examine associations of taxi use and road safety outcomes the way that ridesourcing use relationships have been studied with safety outcomes.

Our analysis aims to bridge literature gaps by leveraging granular local-scale data to shed light on the relationship between transportation network companies' operation and road safety outcomes. In our study, these associations are captured by using real-world ridesourcing trip data. It is crucial to uncover these effects since ridesourcing could be promoted or disincentivized by city managers and policymakers to contribute towards traffic crashes and injuries reduction under a vision zero's plan. Prioritizing successful interventions could be crucial to achieving substantial road safety improvements with ridesourcing services integration.

\section*{Methods}
To measure the association between ridesourcing travel and road safety, we use spatial panel data models that have been previously applied in transportation safety research \cite{Wang2007,Xie2014,AbdelAty2007}. The natural logarithm of crashes, injuries, fatalities, and DWI offenses, as well as ridesourcing exposure, addresses the variables' right skewness via normalization~\cite{Washington2003}. We leverage variables on safety outcomes, ridesourcing and traffic volumes, and socio-demographics from January 2012 to the beginning of April 2017 in our study. To conduct the proposed analysis, we aggregate variables by census tract and month-year units. The data used here (described in greater detail in the next section) are longitudinal, containing repeated observations of the same census tract units over time. Cross-sectional data and models suffer from an inability to capture intertemporal dependence of events, which the panel data models used here are expected to capture. Our proposed models with spatial and time fixed-effects enable reducing bias from unobserved factors that are changing over time but are constant over each spatial unit and controls for unobserved factors that change over space but are constant over time \cite{Washington2003}.

Previous studies that attempt to answer similar research questions with empirical data modeling use the difference-in-differences estimation, which compares how the trajectory of road safety outcomes differs before and after the launch of ridesourcing. Due to the absence of a control group in our study, since the RideAustin operation was launched in Travis County in all census tracts simultaneously (June 2016), we cannot deploy a DiD model. Instead, we employ spatial panel fixed-effects lag and error models and Spatial AutoRegressive with additional AutoRegressive error structure (SARAR) models that allow for the disturbances to be generated by a spatial autoregressive process~\cite{Anselin1995}. The index of each spatial unit is $i \in {{1,…,I}}$, and each time unit is $t \in {{1,…,T}}$. A fixed-effects spatial lag specification is presented in Eq~(\ref{eq:schemeP}), according to~\cite{Anselin1992}:
\begin{eqnarray}
\label{eq:schemeP}
	y_{it}=\lambda \sum_{i\neq j} w_{ij}y_{jt}+\beta x_{it}+\alpha_i+\gamma_t+u_{it},
\end{eqnarray}
where $y_{it}$ is each road safety outcome (e.g., the number of total crashes that occurred in the spatial unit $i$ during month $t$, and similar for injuries, fatalities, and DWI offenses), $w_{ij}$ is a spatial weighted matrix constant over time $t$ with diagonal elements equal to zero in which neighborhood relationships are defined between the spatial units of analysis, $\sum_{i \neq j} w_{ij}y_{jt}$ is the spatially lagged dependent variable which denotes that the value of $y$ at time $t$ is explained not only by the values of exogenous independent variables but also those $y$ neighboring the spatial unit $i$, $\lambda$ is the scalar spatially autoregressive coefficient of the spatially lagged dependent variable, $\alpha_i$ is the spatial unit fixed-effect, $\gamma_t$ is the time unit fixed-effect, $\beta$ the vector of parameters to be estimated, $x_{it}$ a vector of explanatory variables, and the error terms $u_{it} \sim N(\theta,\sigma^2)$. The census tract fixed-effects control for all time-invariant census tract specific factors that are potentially correlated with safety outcomes, such as area. Similar assumptions hold for the time fixed-effects that control for census tract invariant factors that vary by month like travel patterns. For all models, $w_{ij}$ weights are defined based on binary contiguity, where $w_{ij}=1$ when the intersection of the boundaries of $i$ and $j$ spatial units is not empty, otherwise $w_{ij}=0$.
The spatial error with fixed-effects model is presented in Eq~(\ref{eq:scheme2}) and Eq~(\ref{eq:scheme3}), according to the specification in~\cite{Baltagi2003}:
\begin{eqnarray}
\label{eq:scheme2}
	y_{it}=\beta x_{it}+\alpha_i+\gamma_t+u_{it},
\end{eqnarray}
\begin{eqnarray}
\label{eq:scheme3}
	u_{it}=\rho \sum_{i \neq j} w_{ij} u_{jt}+\epsilon_{it},
\end{eqnarray}
where the disturbance term $u_{it}$ follows the first order spatial autoregressive process of the equation presented in Eq~(\ref{eq:scheme3}), and $\rho$ is spatial autoregressive coefficient where $|\rho|<  
1$. The rest of the terms have been specified in the previous paragraph. Comparing the spatial lag and spatial error models, the former suggests a diffusion, where road safety crashes in one spatial unit predict an increased likelihood of road crashes in neighboring places; the latter suggests that we have omitted spatially correlated covariates that would affect inference.

Last, the SARAR model is defined as presented in Eq~(\ref{eq:scheme4}) and Eq~(\ref{eq:scheme5}) from~\cite{Anselin1995,Debarsy2010}:

\begin{eqnarray}
\label{eq:scheme4}
	y_{it}=\lambda \sum_{i \neq j}w_{ij}y_{jt} +\beta x_{it}+\alpha_i+\gamma_t+u_{it},
\end{eqnarray}
\begin{eqnarray}
\label{eq:scheme5}
	u_{it}=\rho \sum_{i\neq j}w_{ij}u_{jt}+\epsilon_{it}, 
\end{eqnarray}
where both $\lambda$ and $\rho$ are spatially autoregressive coefficients. SARAR accounts for both neighboring effects and omitted spatially correlated covariates. Maximization of the likelihood function results in the estimation of the unknown coefficients $\beta$, per the existing notation for the spatial lag and error models~\cite{Anselin1992} and for the SARAR model~\cite{Millo2012}.

We conduct specification tests to confirm that fixed-effects are the most appropriate over random effects. The Hausman test~\cite{Hausman1978} is applied that denotes that a fixed-effects model is at least as consistent as the random-effects specification, and thus, a fixed-effects specification is chosen. Locally robust panel Langrage Multiplier (LM) tests are used to test the absence of each spatial term, for spatial dependence~\cite{Baltagi2003}. The LM statistics tests are used to test for spatial autocorrelation in the form of an endogenous spatial lag variable and spatially autocorrelated errors~\cite{Debarsy2010}.

\section*{Data}

Datasets representative of Travis County, Texas are analyzed including detailed historical road safety measures from the Texas Department of Transportation Crash Records Information System~\cite{TDOT} and DWI offenses from the Austin Police Department Crime Reports~\cite{AustinPD}. For both road safety and DWI offense records, apart from temporal variability, we have access to their exact locations as longitude and latitude coordinates. Ridesourcing use is captured based on actual, comprehensive  trip-level data from RideAustin's open record. RideAustin is a company that operated in Travis County in Texas, without competing with Uber and Lyft between June 2016 and May 2017. RideAustin offered approximately 1.5 million rides~\cite{RideAustin2017} during that period. Note that during the period that Uber and Lyft exited the Austin market, RideAustin was launched and other transportation network companies, such as Fare and Fasten~\cite{Zeitlin2019}. We are not aware of the ridership share that the rest of ridesourcing companies attracted, thus we might underestimate the use of ridesourcing services in the region. However, we know that 47\% of those that used the new transportation network companies in Austin chose RideAustin from a local survey's findings~\cite{Hampshire2017}. RideAustin's market share is sufficient to suggest high representativeness of ridesourcing patterns~\cite{Lavieri2018}.

In our analysis, we aim to control for overall traffic in the region after conducting monthly travel demand analyses using the StreetLight Data platform~\cite{StreetLight}. Normalized trip counts for all census tracts within Travis County during the period of interest are used. The normalized vehicular traffic counts data do not include commercial, ridesourcing, and delivery data according to the specifications of the StreetLight Data platform. The normalized trip counts, used here as a proxy for the travel demand in Austin, are calibrated with annual average daily traffic (AADT) data available at the network level for every link in the Austin metropolitan region~\cite{ACS}. Unfortunately, the shapefiles of the annually variant daily vehicle miles traveled (DVMT) in the Austin region are only suitable for calibrating the StreetLight Data trips counts and not for running the model with DVMT as a control variable. Using annually variant DVMT would result in losing the monthly granularity that we are interested in maintaining across both the ridesourcing use and the overall transportation network use covariates. Additional controls that capture socio-demographic characteristics and their change over the analysis period are available from processed American Community Survey data. The socio-economic data capture demographic and land use variations across census tracts and years. Variables of interest include median household income, population density, employment density, and percent of zero vehicle ownership~\cite{ACS}.

\subsection*{Road safety outcome rates}

\begin{figure}[!h]
\includegraphics[width=\textwidth]{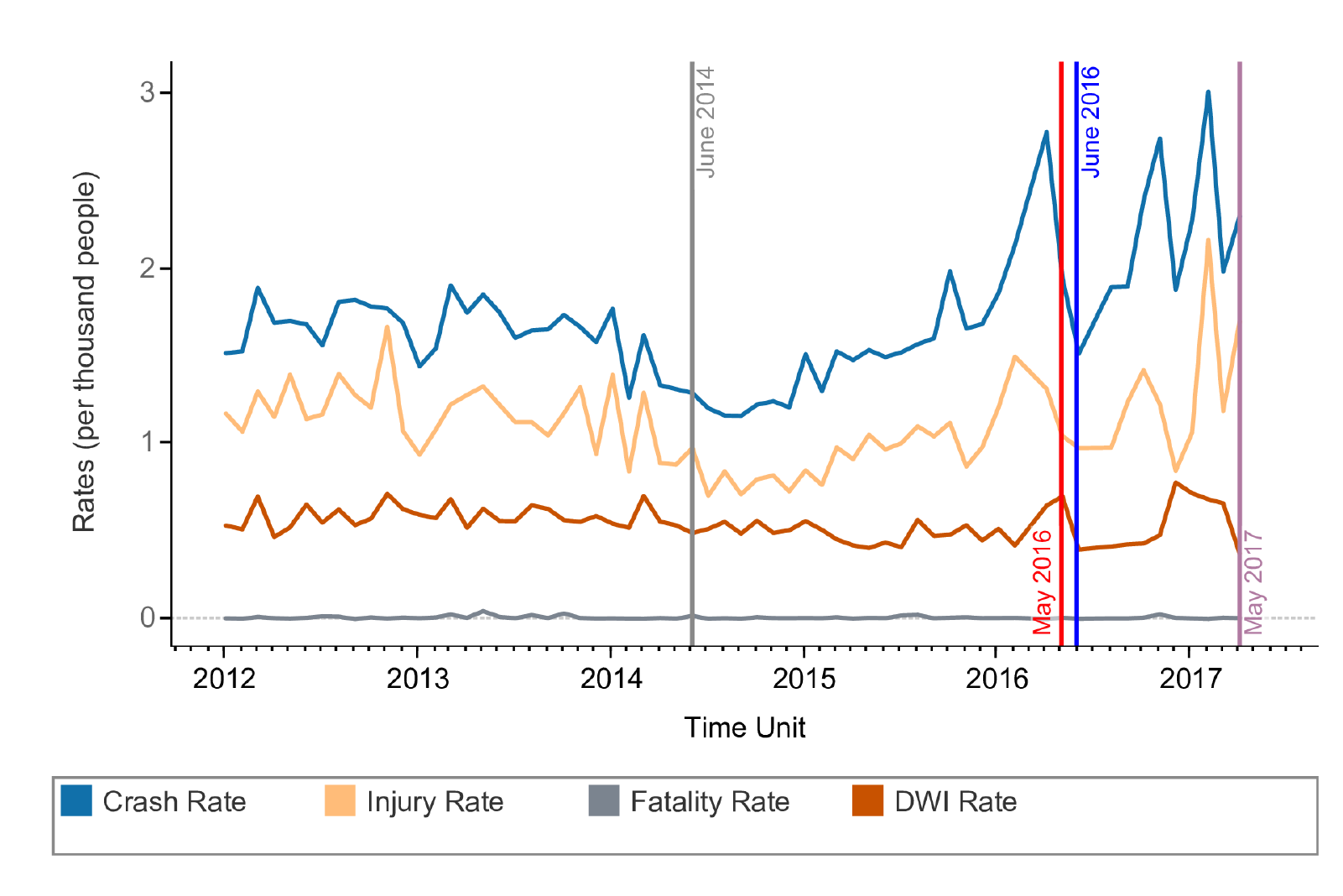}
\caption{{\bf Road safety outcome rates in Travis County, Texas.}}
\label{fig1}
\end{figure}

The traffic safety analysis period is from January 2012 to April 2017. We showcase in Fig~\ref{fig1} the time series of the monthly crash, injury, fatality, and DWI offense rates (per thousand people) averaged by census tract in Travis County. Note that we use the KABCO scale for injury severity~\cite{Klop1999}; the injuries used to compute the corresponding rate include incapacitating (A), non-incapacitating (B), and possible injuries indicated by behavior but not visible wounds (C)~\cite{FHWA}. We observe similarities between the crash and injury rates seasonal trends, even though for the former the fitted linear trend over the total period of time is increasing and for the latter decreasing. 

In Fig~\ref{fig1}, we overlay the monthly ridesourcing operation timeline in Travis County with traffic safety outcome rates. The period between January 2012 and June 2014 is characterized by no ridesourcing operation. Uber and Lyft launched in the region during June 2014 and operated there until May 9, 2016, when they stopped offering services due to the city of Austin's regulation that required fingerprint background checks on their drivers~\cite{Solomon2017}. At the time of Uber/Lyft entry, crashes and injuries rates were declining; while operating and before Uber's and Lyft's exit, crash and injury rates were rising. Other effects might play a role in that, such as the city's population growth and the significant gasoline price drops in 2014, which are positively associated with additional travel demand. RideAustin, a ridesourcing service, was launched in Travis County in June 2016 and operated until May 29, 2017, when Uber and Lyft returned. Due to the unavailability of ridesourcing use data from these major transportation network companies between June 2014 and May 2016, we exclude that period from our analysis. The group of observations before ridesourcing entry in June 2014 until the end of May 2014 are denoted as a "before ridesourcing" group and the observations group after RideAustin's launch in June 2016 up to the beginning of April 2017 as the "after ridesourcing" group. 

\subsection*{Ridesourcing use}

\begin{figure}[!h]
\includegraphics[width=\textwidth]{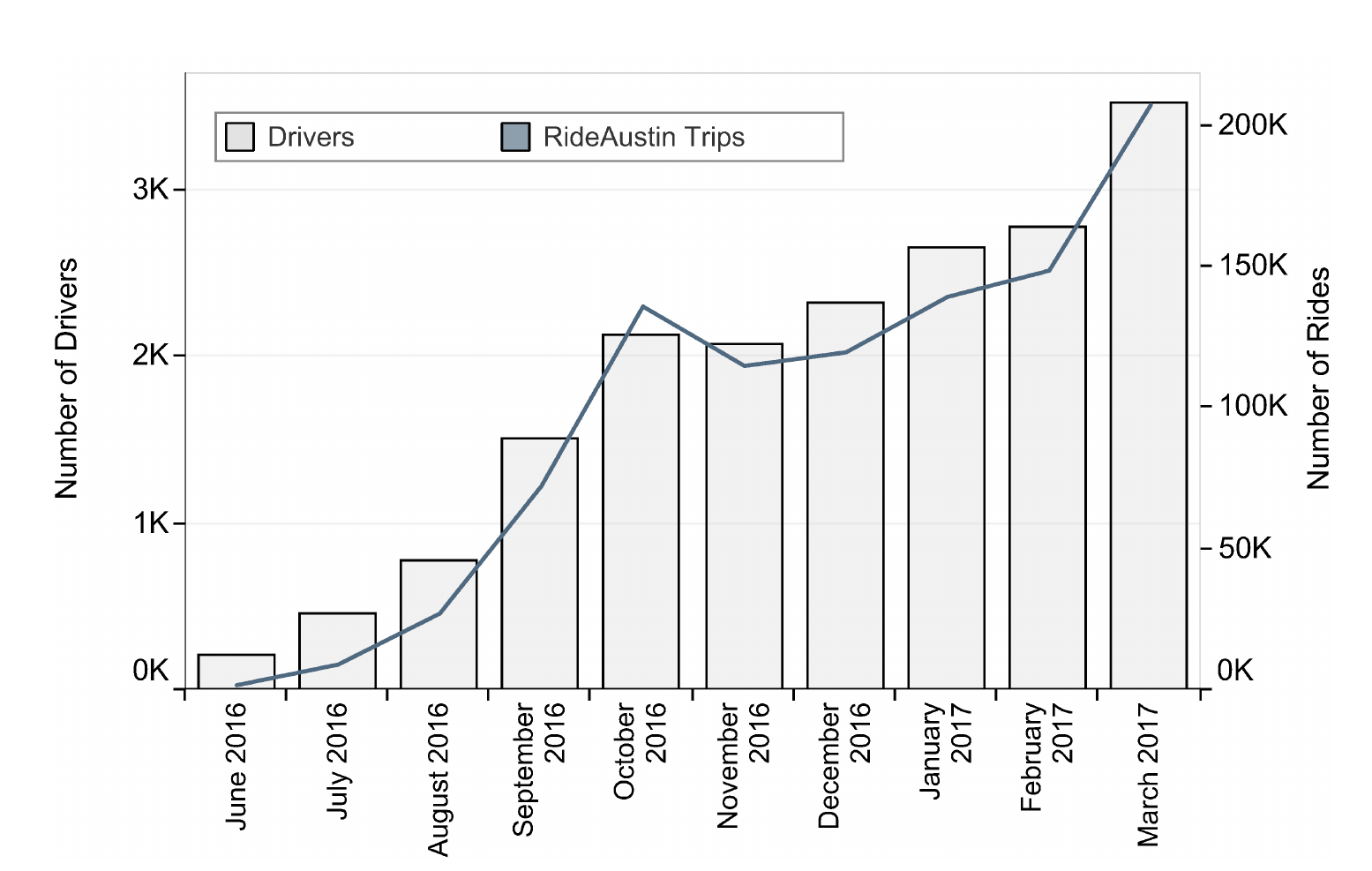}
\caption{{\bf Ridesourcing use in Travis County, Texas.}}
\label{fig2}
\end{figure}

The monthly timeline presenting the number of RideAustin trips that were conducted in Travis County is presented in Fig~\ref{fig2}, using RideAustin data~\cite{RideAustin2017}. Specifically, the line graph presents the monthly number of rides given (corresponding right y-axis), and the bar graph shows the monthly number of drivers that offered rides (corresponding left y-axis). The number of rides offered between RideAustin entry's and October 2016 is increasing exponentially; then, the number of rides keeps increasing at a lower rate, reaching a peak during the South by Southwest festival and conference season. Partial trip data for a few days in April 2017 are available; we truncate the dataset to include ridesourcing trip origin and destination (OD) counts only up to the beginning of April 2017. 

\begin{figure}[!h]
\includegraphics[width=.5\textwidth]{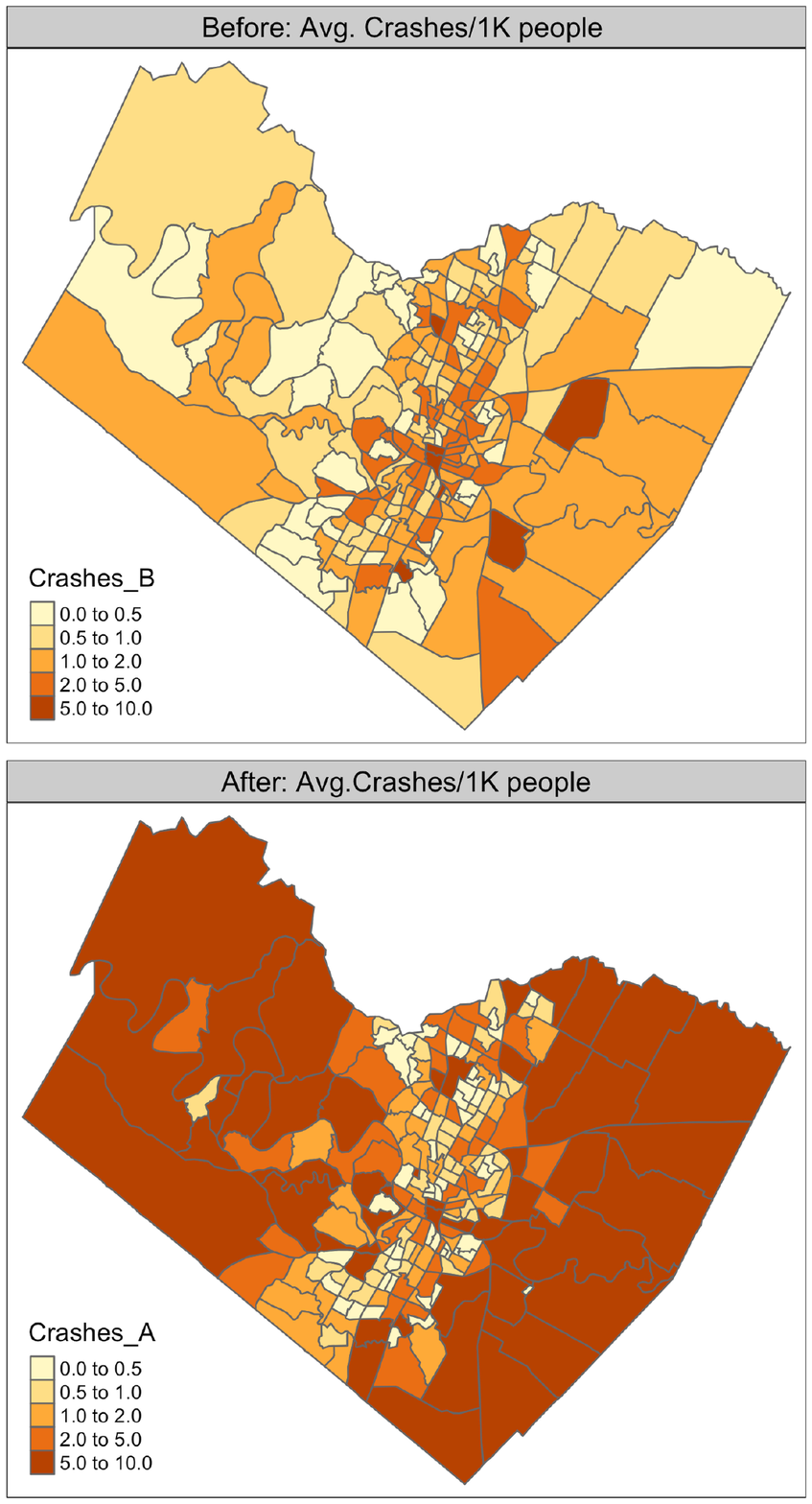}\hfill
\includegraphics[width=.5\textwidth]{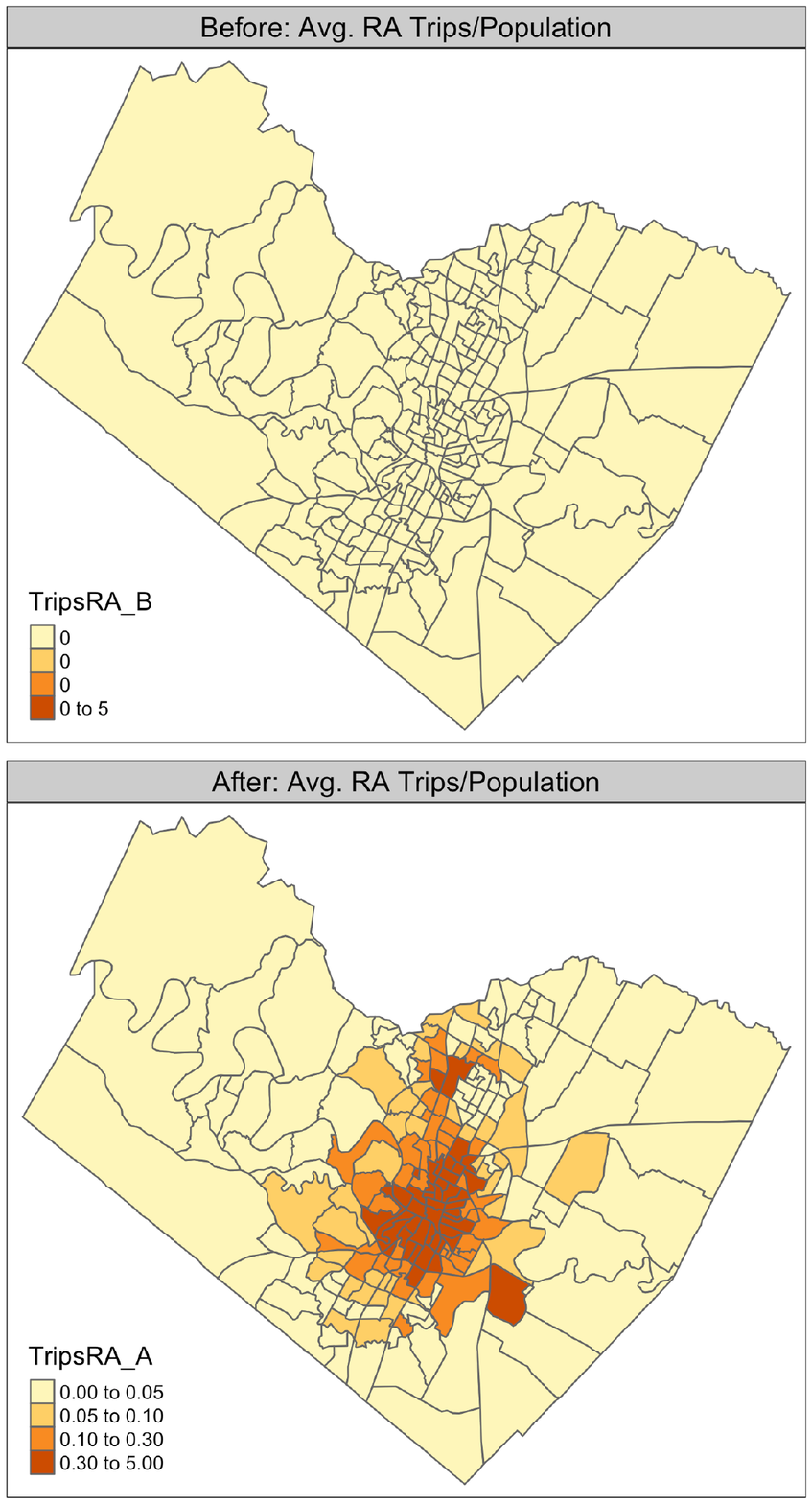}
\caption{{\bf Road crash rate and ridesourcing rate before and after the introduction of the ridesourcing service in Travis County, Texas census tracts.}
Left panel: Average crashes per 1000 people. Right panel: Average RideAustin trips per census tract population.}
\label{fig3}
\end{figure}

The average monthly crash rate and RideAustin passengers' pick-up and drop-off rate for each census tract in Travis County are portrayed in Fig~\ref{fig3}. We are grouping together ridesourcing trips origins and destinations that fall in our spatial units of analysis (i.e., census tracts), without double-counting trips that have both their origin and destination in the same census tract. Due to this grouping we might not be able to capture that loading and dropping off of customers might have different impacts on the congestion across census tracts and can result in traffic stress levels. We present these indicators in Fig~\ref{fig3}, denoting their shares before and after the entry of ridesourcing in the region, without accounting for the excluded period shown in Fig~\ref{fig1}. In June 2016, RideAustin started offering rides mainly in the downtown Austin region (smaller area census tracts in the center of the map) and then, over the months, expanded their service radius to cover all census tracts within the Travis County. 

With highway networks spanning the suburban Austin region, we observe an increase of average crashes per 1000 people in the aftermath of RideAustin launch there. Fig~\ref{fig3} also suggests that after RideAustin's entry, the rate of total crashes increased in the suburbs of Travis County but not in the downtown regions that the ridesourcing operation covered at a higher rate. The crashes increase in the suburbs could be associated with population and firm growth and could reflect changes in travel patterns that resulted from spatial differences due to the region's economic development. 

\subsection*{Other controls}

Various measures are introduced as control variables in our model. Their selection is based on prior evidence (as shown in Table~\ref{table1}) and includes median household income~\cite{Greenwood2017,Barrios2019}, percent of zero vehicle ownership~\cite{Barrios2019}, population density~\cite{Greenwood2017,Barrios2019,Dills2018}, and percentage of employment~\cite{Barrios2019,Brazil2016,MartinBuck,Dills2018}. All these covariates are available through the American Community Survey for every census tract on an annual basis~\cite{ACS}. We make the assumption that such socio-demographic changes would be mainly observed on an annual basis since decisions of household relocation are made on an annual basis (e.g.,~\cite{Clark2017}). Given the land use changes in Travis County the latest years, we control for the monthly spatial variation in trip OD counts per spatial-time unit of analysis using normalized trip count data from the StreetLight Data platform~\cite{StreetLight}. The normalized OD trips, derived from the underlying data sample per census tract per month, is available as an output of the StreetLight Data travel demand modeling. Unfortunately, we only have the ability to estimate traffic exposure from January 2016 to December 2018 within the StreetLight Data platform due to the unavailability of historical datasets. This period slightly overlaps with our analysis period. Time series analysis is performed to estimate traffic exposure historical data (for more information, see \nameref{S1_Appendix}). 

The hypotheses that support the inclusion of the aforementioned control variables are as follows. Population density, employment percentage, and traffic (provided through the StreetLight Data's platform) are expected to capture the market's size. The median household income is controlling the market's wealth~\cite{Greenwood2017}. Traffic crashes are significantly increased in cities with lower average per capita income~\cite{Barrios2019}. The percentage of zero vehicle ownership is expected to capture likely users who do not own a personal automobile and may rely on other modes of transport~\cite{Hall2018}. Note that, commonly, when examining factors contributing to crash risk, exposure measures tend to include roadway length~\cite{Huang2010}. In our case, the number of roadway miles does not vary at the monthly level per each spatio-temporal unit of analysis. Thus, any unobserved effects can be captured with the spatial effects' parameter entered in our proposed models.  

Table~\ref{table2} presents descriptive statistics for our sample of 8,720 census tract-month-year units, corresponding to 218 different census tracts in Travis County and 40 monthly periods. The mean, standard deviation, and median values of the safety outcomes, as well as the rest of the socio-economic, travel, and transportation specific covariates are presented there. Only road crashes, fatalities, and injuries that involve light-duty vehicles are considered in the analysis (excluding crashes that only involve trucks, buses, and trains).

\begin{table}[!ht]
\begin{adjustwidth}{-2.25in}{0in}
\centering
\caption{
{\bf Summary statistics of census tract-month-year units of analysis.}}
\begin{tabular}{|l|l|l|l|l|l|}
\hline
& \textbf{Mean} & \textbf{Std. Dev.} & \textbf{Median} & \textbf{Minimum} & \textbf{Maximum} \\ \hline
\thickhline
\textbf{Crashes}                           & 6.16   & 5.65               & 5.00   & 0.00    & 49.00   \\ \hline
\textbf{Fatalities}                        & 0.041  & 0.22               & 0.00   & 0.00    & 3.00    \\ \hline
\textbf{Injuries}                          & 4.13   & 4.70               & 3.00   & 0.00    & 75.00   \\ \hline
\textbf{DWI offenses}                      & 2.09   & 4.29               & 1.00   & 0.00    & 66.00   \\ \hline
\textbf{Ridesourcing trips}                & 336.5  & 3,041.13           & 0.00   & 0.00    & 163,788 \\ \hline
\textbf{Median household (hh) income (\$)}        & 65,757 & 33,387             & 58,539 & 0       & 216,875 \\ \hline
\textbf{OD trips}                          & 27,507 & 22,941             & 20,703 & 12,405  & 219,626 \\ \hline
\textbf{Gas price (2018 \$/gallon)}        & 3.00   & 0.61               & 3.23   & 1.94    & 3.78    \\ \hline
\textbf{Population density (pop /sq ft)}   & 44,923 & 41,373             & 40,004 & 0       & 293,348 \\ \hline
\textbf{Percent of zero vehicle ownership} & 3.2\%  & 3.7\%              & 2.0\%  & 0.0\%   & 25.3\%  \\ \hline
\textbf{Percent of employment}             & 72.0\% & 11.6\%             & 73.9\% & 0.0\%   & 94.1\%  \\ \hline
\textbf{Records}                           & \multicolumn{5}{l|}{8,720} \\ \hline     
\end{tabular}
\label{table2}
\end{adjustwidth}
\end{table}

\section*{Results and discussion}
Descriptive statistics of our sample of 8,720 census-tract month units, categorized by ridesourcing presence in Travis County, are presented in Table~\ref{table3}. Ridesourcing was operating in 27.50\% of the total census tract-month units of our analysis. Crash and injury rates are higher during the months when ridesourcing was offered. On the contrary, DWI offenses rate and fatality rates are lower, even though the averages' difference is much smaller. We observe an increase in the normalized trip counts rate per census tract and month unit, accompanied by an increase of the median household income and a significant decrease in gas prices, as well as an increase of the average population in the region, after the ridesourcing entry.

\begin{table}[!ht]
\begin{adjustwidth}{-2.25in}{0in}
\centering
\caption{
{\bf Descriptive Statistics: Safety and other metrics magnitude before and after ridesourcing entry periods.}}
\begin{tabular}{|wl{6.6cm}|l|l|}
\hline
                                                                & \textbf{Before Ridesourcing} & \textbf{After Ridesourcing} \\ 
                                                & \textbf{Jan 2012-May 2014}   & \textbf{Jun 2016-Apr 2017}  \\ 
                                                                & Mean (Std. Dev.)             & Mean (Std. Dev.)            \\ \hline
\thickhline
\textbf{Crash rate}                                             & 1.65 (3.22)                  & 2.20 (11.54)                \\ \hline
\textbf{DWI offense rate}                                       & 0.59 (1.35)                  & 0.53 (2.56)                 \\ \hline
\textbf{Injury rate}                                            & 1.17 (2.97)                  & 1.38 (9.93)                 \\ \hline
\textbf{Fatality rate}                                          & 0.014 (0.17)                 & 0.010 (0.07)                \\ \hline
\textbf{Ridesourcing rate}                                      & n.a.                         & 0.28 (1.05)                 \\ \hline
\multirow{2}{*}{\textbf{Median hh income (\$)}}             & 63,744                       & 71,068                      \\
                                                                & (32,930)                     & (34,002)                    \\ \hline
\multirow{2}{*}{\textbf{OD trips rate (trips/population)}}      & 6.27                         & 7.60                        \\
                                                                & (8.31)                       & (14.91)                     \\ \hline
\multirow{2}{*}{\textbf{Gas price (2018 \$/gallon)}}             & 3.36                         & 2.00                        \\
                                                                & (0.19)                       & (0.08)                      \\ \hline
\multirow{2}{*}{\textbf{Census tract population}}               & 4,847                        & 5,314                       \\
                                                                & (2,620)                      & (3,029)                     \\ \hline
\multirow{2}{*}{\textbf{Percent of zero vehicle ownership (\%)}} & 3.33\%                        & 3.02\%                       \\
                                                                & (3.78\%)                      & (3.47\%)                     \\ \hline
\multirow{2}{*}{\textbf{Percent of employment (\%)}}                  & 72.03\%                       & 72.10\%                      \\
                                                                & (11.78\%)                     & (10.93\%)                    \\ \hline
\textbf{Records (\%)}                                            & 72.50\%                       & 27.50\%                      \\ \hline      
\end{tabular}
\begin{flushleft}Rates covariates are measured per 1,000 people (based on the population variable of the American Community Survey, which is annually available for each census tract spatial unit)~\cite{ACS}.
\end{flushleft}
\label{table3}
\end{adjustwidth}
\end{table}

The spatial fixed-effects panel data model estimates, with all controls included, are presented in Table~\ref{table4}, Table~\ref{table5}, and Table~\ref{table6} for the spatial lag, spatial error, and SARAR models, respectively. Lagrange multiplier tests results are provided there. Signs and magnitudes of the $\beta$ coefficients are consistent across the three models, supporting the robustness of our findings.

\begin{table}[!ht]
\begin{adjustwidth}{-2.25in}{0in}
\centering
\caption{
{\bf Spatial fixed-effects panel data modeling results: spatial lag.}}
\begin{tabular}{|l|ll|ll|ll|ll|}
\hline
                                                   & \multicolumn{2}{l|}{\textbf{Log(1+Crashes)}}        & \multicolumn{2}{l|}{\textbf{Log(1+Injuries)}}      & \multicolumn{2}{l|}{\textbf{Log(1+Fatalities)}}    & \multicolumn{2}{l|}{\textbf{Log(1+DWI)}}            \\
                                                   & \textbf{$\beta$}                     & \textbf{}    & \textbf{$\beta$}                     & \textbf{}   & \textbf{$\beta$}                     & \textbf{}   & \textbf{$\beta$}                     & \textbf{}    \\ \hline
\thickhline
\multirow{2}{*}{\textbf{Percent of employment}}             & \multicolumn{1}{l|}{-0.009}          &              & \multicolumn{1}{l|}{0.337}           &             & \multicolumn{1}{l|}{-0.111}          & \textbf{*}  & \multicolumn{1}{l|}{0.036}           &              \\
                                                   & \multicolumn{1}{l|}{{[}0.165{]}}     &              & \multicolumn{1}{l|}{{[}0.222{]}}     &             & \multicolumn{1}{l|}{{[}0.052{]}}     &             & \multicolumn{1}{l|}{{[}0.159{]}}     &              \\ \hline
\multirow{2}{*}{\textbf{Median HH income}}                  & \multicolumn{1}{l|}{$-1.75 10^{-6}$}      & \textbf{.}   & \multicolumn{1}{l|}{$-1.07 10^{-6}$}      &             & \multicolumn{1}{l|}{$0.56 10^{-6}$}       & \textbf{.}  & \multicolumn{1}{l|}{$-0.19 10^{-6}$}      & \textbf{.}   \\
                                                   & \multicolumn{1}{l|}{{[}$1.05 10^{-6}${]}} &              & \multicolumn{1}{l|}{{[}$1.41 10^{-6}${]}} &             & \multicolumn{1}{l|}{{[}$0.33 10^{-6}${]}} &             & \multicolumn{1}{l|}{{[}$1.01 10^{-6}${]}} &              \\ \hline
\multirow{2}{*}{\textbf{Percent of zero vehicle ownership}} & \multicolumn{1}{l|}{-0.837}          & \textbf{**}  & \multicolumn{1}{l|}{-0.851}          & \textbf{*}  & \multicolumn{1}{l|}{0.116}           &             & \multicolumn{1}{l|}{-0.0862}         &              \\
                                                   & \multicolumn{1}{l|}{{[}0.309{]}}     &              & \multicolumn{1}{l|}{{[}0.417{]}}     &             & \multicolumn{1}{l|}{{[}0.097{]}}     &             & \multicolumn{1}{l|}{{[}0.298{]}}     &              \\ \hline
\textbf{Population density}                                 & \multicolumn{1}{l|}{$3.65 10^{-6}$}       & \textbf{**}  & \multicolumn{1}{l|}{$2.56 10^{-6}$}       &             & \multicolumn{1}{l|}{$-1.23 10^{-6}$}      & \textbf{**} & \multicolumn{1}{l|}{$1.70 10^{-6}$}       &              \\
                                                   & \multicolumn{1}{l|}{{[}$1.33 10^{-6}${]}} &              & \multicolumn{1}{l|}{{[}$1.79 10^{-6}${]}} &             & \multicolumn{1}{l|}{{[}$0.42 10^{-6}${]}} &             & \multicolumn{1}{l|}{{[}$0.13 10^{-6}${]}} &              \\ \hline
\textbf{OD trips}                                           & \multicolumn{1}{l|}{$3.35 10^{-8}$}       &              & \multicolumn{1}{l|}{$0.86 10^{-6}$}       &             & \multicolumn{1}{l|}{$0.13 10^{-6}$}       &             & \multicolumn{1}{l|}{$1.23 10^{-6}$}       & \textbf{*}   \\
                                                   & \multicolumn{1}{l|}{{[}$6.24 10^{-7}${]}} &              & \multicolumn{1}{l|}{{[}$0.84 10^{-6}${]}} &             & \multicolumn{1}{l|}{{[}$0.19 10^{-6}${]}} &             & \multicolumn{1}{l|}{{[}$0.60 10^{-6}${]}} &              \\ \hline
\textbf{Log(1+Trips RideAustin)}                            & \multicolumn{1}{l|}{-0.013}          & \textbf{.}   & \multicolumn{1}{l|}{-0.026}          & \textbf{**} & \multicolumn{1}{l|}{-0.0009}         &             & \multicolumn{1}{l|}{-0.0372}         & \textbf{***} \\
                                                   & \multicolumn{1}{l|}{{[}0.007{]}}     &              & \multicolumn{1}{l|}{{[}0.009{]}}     &             & \multicolumn{1}{l|}{{[}0.002{]}}     &             & \multicolumn{1}{l|}{{[}0.007{]}}     &              \\ \hline
\multirow{2}{*}{\textbf{$\lambda$}}                         & \multicolumn{1}{l|}{0.114}           & \textbf{***} & \multicolumn{1}{l|}{0.055}           & \textbf{**} & \multicolumn{1}{l|}{0.003}           &             & \multicolumn{1}{l|}{0.05}            & \textbf{**}  \\
                                                   & \multicolumn{1}{l|}{{[}0.018{]}}     & \textbf{}    & \multicolumn{1}{l|}{{[}0.018{]}}     & \textbf{}   & \multicolumn{1}{l|}{{[}0.018{]}}     &             & \multicolumn{1}{l|}{{[}0.018{]}}     & \textbf{}    \\ \hline
\textbf{LM test (df=1)}                                     & \multicolumn{1}{l|}{42.36}           & \textbf{***} & \multicolumn{1}{l|}{9.43}            & \textbf{.}  & \multicolumn{1}{l|}{0.03}            & \textbf{}   & \multicolumn{1}{l|}{7.56}            & \textbf{.}   \\ \hline

\end{tabular}
\label{table4}
\begin{flushleft}Symbol \textbf{***} corresponds to p$<0.0001$, \textbf{**} to p$<0.001$, \textbf{*} to p$<0.01$, and \textbf{.} to p$<0.05$.
\end{flushleft}
\end{adjustwidth}
\end{table}

\begin{table}[!ht]
\begin{adjustwidth}{-2.25in}{0in}
\centering
\caption{
{\bf Spatial fixed-effects panel data modeling results: error lag.}}
\begin{tabular}{|l|ll|ll|ll|ll|}
\hline
                                       & \multicolumn{2}{l|}{\textbf{Log(1+Crashes)}}        & \multicolumn{2}{l|}{\textbf{Log(1+Injuries)}}       & \multicolumn{2}{l|}{\textbf{Log(1+Fatalities)}}     & \multicolumn{2}{l|}{\textbf{Log(1+DWI)}}            \\
                                       & \textbf{$\beta$}                     & \textbf{}    & \textbf{$\beta$}                     & \textbf{}   & \textbf{$\beta$}                      & \textbf{}   & \textbf{$\beta$}                     & \textbf{}    \\ \hline
\thickhline
\multirow{2}{*}{\textbf{Percent of employment}} & \multicolumn{1}{l|}{0.004}           &              & \multicolumn{1}{l|}{0.342}           &             & \multicolumn{1}{l|}{-0.111}           & \textbf{*}  & \multicolumn{1}{l|}{0.0378}          &              \\
                                       & \multicolumn{1}{l|}{{[}0.165{]}}     &              & \multicolumn{1}{l|}{{[}0.222{]}}     &             & \multicolumn{1}{l|}{{[}0.052{]}}      &             & \multicolumn{1}{l|}{{[}0.1589{]}}    &              \\ \hline
\textbf{Median HH income}                       & \multicolumn{1}{l|}{$-1.70 10^{-6}$}      &              & \multicolumn{1}{l|}{$-1.05 10^{-6}$}      &             & \multicolumn{1}{l|}{$0.56 10^{-6}$}        & \textbf{.}  & \multicolumn{1}{l|}{$-1.85 10^{-6}$}      & \textbf{.}   \\
                                       & \multicolumn{1}{l|}{{[}$1.05 10^{-6}${]}} &              & \multicolumn{1}{l|}{{[}$1.41 10^{-6}${]}} &             & \multicolumn{1}{l|}{{[}$0.33 10^{-7}${]}}  &             & \multicolumn{1}{l|}{{[}$1.00 10^{-6}${]}} &              \\ \hline
\textbf{Percent of zero vehicle ownership}      & \multicolumn{1}{l|}{$-0.785$}          & \textbf{*}   & \multicolumn{1}{l|}{-0.829}          & \textbf{*}  & \multicolumn{1}{l|}{0.116}            &             & \multicolumn{1}{l|}{-0.077}          &              \\
                                       & \multicolumn{1}{l|}{{[}0.312{]}}     &              & \multicolumn{1}{l|}{{[}0.419{]}}     &             & \multicolumn{1}{l|}{{[}0.098{]}}      &             & \multicolumn{1}{l|}{{[}0.299{]}}     &              \\ \hline
\textbf{Population density}                     & \multicolumn{1}{l|}{$3.51 10^{-6}$}       & \textbf{**}  & \multicolumn{1}{l|}{$2.44 10-6$}       &             & \multicolumn{1}{l|}{$-1.23 10^{-6}$}       & \textbf{**} & \multicolumn{1}{l|}{$1.68 10^{-6}$}       &              \\
                                       & \multicolumn{1}{l|}{{[}$1.32 10^{-6}${]}} &              & \multicolumn{1}{l|}{{[}$1.79 10^{-6}${]}} &             & \multicolumn{1}{l|}{{[}$0.42 10^{-6}${]}}  &             & \multicolumn{1}{l|}{{[}$1.28 10^{-6}${]}} &              \\ \hline
\textbf{OD trips}                              & \multicolumn{1}{l|}{$7.61 10^{-8}$}       &              & \multicolumn{1}{l|}{$8.65 10^{-7}$}       &             & \multicolumn{1}{l|}{$0.13 10^{-6}$}        &             & \multicolumn{1}{l|}{$1.28 10^{-6}$}       & \textbf{*}   \\
                                       & \multicolumn{1}{l|}{{[}$6.44 10^{-7}${]}} &              & \multicolumn{1}{l|}{{[}$8.55 10^{-7}${]}} &             & \multicolumn{1}{l|}{{[}$0.198 10^{-6}${]}} &             & \multicolumn{1}{l|}{{[}$0.61 10^{-6}${]}} &              \\ \hline
\textbf{Log (1+Trips RideAustin)}                & \multicolumn{1}{l|}{-0.0133}         & \textbf{.}   & \multicolumn{1}{l|}{-0.026}          & \textbf{**} & \multicolumn{1}{l|}{-0.0008}          &             & \multicolumn{1}{l|}{-0.0381}         & \textbf{***} \\
                                       & \multicolumn{1}{l|}{{[}0.007{]}}     &              & \multicolumn{1}{l|}{{[}0.009{]}}     &             & \multicolumn{1}{l|}{{[}0.002{]}}      &             & \multicolumn{1}{l|}{{[}0.007{]}}     &              \\ \hline
\multirow{2}{*}{\textbf{$\rho$}}                & \multicolumn{1}{l|}{0.112}           & \textbf{***} & \multicolumn{1}{l|}{0.112}           & \textbf{**} & \multicolumn{1}{l|}{0.0038}           &             & \multicolumn{1}{l|}{0.0473}          & \textbf{**}  \\
                                       & \multicolumn{1}{l|}{{[}0.018{]}}     & \textbf{}    & \multicolumn{1}{l|}{{[}0.018{]}}     & \textbf{}   & \multicolumn{1}{l|}{{[}0.0183{]}}     & \textbf{}   & \multicolumn{1}{l|}{{[}0.0181{]}}    & \textbf{}    \\ \hline
\textbf{LM test (df=1)}                         & \multicolumn{1}{l|}{40.01}           & \textbf{***} & \multicolumn{1}{l|}{8.95}            & \textbf{.}  & \multicolumn{1}{l|}{0.04}             & \textbf{}   & \multicolumn{1}{l|}{6.68}            & \textbf{**}  \\ \hline

\end{tabular}
\label{table5}
\begin{flushleft}Symbol \textbf{***} corresponds to p$<0.0001$, \textbf{**} to p$<0.001$, \textbf{*} to p$<0.01$, and \textbf{.} to p$<0.05$.
\end{flushleft}
\end{adjustwidth}
\end{table}

\begin{table}[!ht]
\begin{adjustwidth}{-2.25in}{0in}
\centering
\caption{
{\bf Spatial fixed-effects panel data modeling results: SARAR.}}
\begin{tabular}{|l|ll|ll|ll|ll|}
\hline
                                  & \multicolumn{2}{l|}{\textbf{Log(1+Crashes)}}        & \multicolumn{2}{l|}{\textbf{Log(1+Injuries)}}       & \multicolumn{2}{l|}{\textbf{Log(1+Fatalities)}}     & \multicolumn{2}{l|}{\textbf{Log(1+DWI)}}            \\
                                  & \textbf{$\beta$}                     & \textbf{}    & \textbf{$\beta$}                     & \textbf{}    & \textbf{$\beta$}                     & \textbf{}    & \textbf{$\beta$}                     & \textbf{}    \\ \hline
\thickhline
\textbf{Percent of employment}    & \multicolumn{1}{l|}{-0.029}          &              & \multicolumn{1}{l|}{0.326}           &              & \multicolumn{1}{l|}{-0.108}          & \textbf{*}   & \multicolumn{1}{l|}{0.034}           &              \\
                                  & \multicolumn{1}{l|}{{[}0.159{]}}     &              & \multicolumn{1}{l|}{{[}0.2024{]}}    &              & \multicolumn{1}{l|}{{[}0.0515{]}}    &              & \multicolumn{1}{l|}{{[}0.158{]}}     &              \\ \hline
\textbf{Median HH income}         & \multicolumn{1}{l|}{$-1.80 10^{-6}$}      & \textbf{.}   & \multicolumn{1}{l|}{$-1.13 10^{-6}$}      &              & \multicolumn{1}{l|}{$0.59 10^{-6}$}       & \textbf{.}   & \multicolumn{1}{l|}{$-1.97 10^{-6}$}      & \textbf{*}   \\
                                  & \multicolumn{1}{l|}{{[}$1.02 10^{-6}${]}} &              & \multicolumn{1}{l|}{{[}$1.40 10^{-6}${]}} &              & \multicolumn{1}{l|}{{[}$0.33 10^{-6}${]}} &              & \multicolumn{1}{l|}{{[}$1.00 10^{-6}${]}} &              \\ \hline
\textbf{Percent of zero vehicle ownership} & \multicolumn{1}{l|}{-0.922}          & \textbf{**}  & \multicolumn{1}{l|}{-0.899}          & \textbf{*}   & \multicolumn{1}{l|}{0.121}           &              & \multicolumn{1}{l|}{-0.103}          &              \\
                                  & \multicolumn{1}{l|}{{[}0.293{]}}     &              & \multicolumn{1}{l|}{{[}0.410{]}}     &              & \multicolumn{1}{l|}{{[}0.099{]}}     &              & \multicolumn{1}{l|}{{[}0.294{]}}     &              \\ \hline
\textbf{Population density}                & \multicolumn{1}{l|}{$3.89 10{-6}$}       & \textbf{**}  & \multicolumn{1}{l|}{$2.86 10^{-6}$}       &              & \multicolumn{1}{l|}{$-1.23 10^{-6}$}      & \textbf{**}  & \multicolumn{1}{l|}{$1.73 10^{-6}$}       &              \\
                                  & \multicolumn{1}{l|}{{[}$1.46 10^{-6}${]}} &              & \multicolumn{1}{l|}{{[}$1.78 10^{-6}${]}} &              & \multicolumn{1}{l|}{{[}$0.41 10^{-6}${]}} &              & \multicolumn{1}{l|}{{[}$1.27 10^{-6}${]}} &              \\ \hline
\textbf{OD trips}                          & \multicolumn{1}{l|}{$1.13 10^{-8}$}       &              & \multicolumn{1}{l|}{$8.55 10^{-7}$}       &              & \multicolumn{1}{l|}{$0.135 10^{-6}$}      &              & \multicolumn{1}{l|}{$1.13 10^{-6}$}       & \textbf{.}   \\
                                  & \multicolumn{1}{l|}{{[}$5.60 10^{-7}${]}} &              & \multicolumn{1}{l|}{{[}$8.04 10^{-7}${]}} &              & \multicolumn{1}{l|}{{[}$0.21 10^{-6}${]}} &              & \multicolumn{1}{l|}{{[}$0.59 10^{-6}${]}} &              \\ \hline
\textbf{Log(1+Trips RideAustin)}           & \multicolumn{1}{l|}{-0.011}          & \textbf{.}   & \multicolumn{1}{l|}{-0.024}          & \textbf{**}  & \multicolumn{1}{l|}{-0.001}          &              & \multicolumn{1}{l|}{-0.036}          & \textbf{***} \\
                                  & \multicolumn{1}{l|}{{[}0.006{]}}     &              & \multicolumn{1}{l|}{{[}0.009{]}}     &              & \multicolumn{1}{l|}{{[}0.003{]}}     &              & \multicolumn{1}{l|}{{[}0.008{]}}     &              \\ \hline
\textbf{$\lambda$}                         & \multicolumn{1}{l|}{0.359}           & \textbf{***} & \multicolumn{1}{l|}{0.186}           &              & \multicolumn{1}{l|}{-0.296}          & \textbf{***} & \multicolumn{1}{l|}{0.136}           &              \\
                                  & \multicolumn{1}{l|}{{[}0.058{]}}     &              & \multicolumn{1}{l|}{{[}0.129{]}}     &              & \multicolumn{1}{l|}{{[}0.089{]}}     &              & \multicolumn{1}{l|}{{[}0.170{]}}     &              \\ \hline
\textbf{$\rho$}                            & \multicolumn{1}{l|}{-0.29}           & \textbf{***} & \multicolumn{1}{l|}{-0.141}          &              & \multicolumn{1}{l|}{0.272}           & \textbf{***} & \multicolumn{1}{l|}{-0.092}          &              \\
                                  & \multicolumn{1}{l|}{{[}0.075{]}}     & \textbf{}    & \multicolumn{1}{l|}{{[}0.146{]}}     & \textbf{}    & \multicolumn{1}{l|}{{[}0.073{]}}     & \textbf{}    & \multicolumn{1}{l|}{{[}0.186{]}}     & \textbf{}    \\ \hline
\textbf{LM: lag (df=1)}                    & \multicolumn{1}{l|}{28.09}           & \textbf{***} & \multicolumn{1}{l|}{6.67}            & \textbf{*}   & \multicolumn{1}{l|}{1.26}            & \textbf{}    & \multicolumn{1}{l|}{11.23}           & \textbf{*}   \\ \hline
\textbf{LM: error (df=1)}                  & \multicolumn{1}{l|}{25.74}           & \textbf{***} & \multicolumn{1}{l|}{6.203}           & \textbf{.}   & \multicolumn{1}{l|}{1.27}            & \textbf{}    & \multicolumn{1}{l|}{10.36}           & \textbf{.}   \\ \hline
\textbf{Hausman test (df=6) chi-squared}   & \multicolumn{1}{l|}{205.6}           & \textbf{***} & \multicolumn{1}{l|}{110.31}          & \textbf{***} & \multicolumn{1}{l|}{10.32}           & .            & \multicolumn{1}{l|}{158.68}          & \textbf{*}   \\ \hline

\end{tabular}
\label{table6}
\begin{flushleft}Symbol \textbf{***} corresponds to p$<0.0001$, \textbf{**} to p$<0.001$, \textbf{*} to p$<0.01$, and \textbf{.} to p$<0.05$.
\end{flushleft}
\end{adjustwidth}
\end{table}

Ridesourcing use is found significantly associated with three road safety outcomes: crashes, injuries, and DWI offenses. Based on the SARAR model coefficients shown in Table~\ref{table6}, for a 10\% increase in ridesourcing use, we expect a 0.12\% decrease in road crashes (p$<0.05$), a 0.25\% decrease in road injuries (p$<0.001$), and a 0.36\% decrease in DWI offenses (p$<0.0001$). Ridesourcing use is not found associated with fatalities at the 0.1 significance level. These results are well aligned with Dills and Mulholland and Morrison et al. findings that associate the entry of ridesourcing with a decrease in DWI offenses and alcohol-involved crashes, respectively~\cite{Morrison2018,Dills2018}. However, Morrison et al. find no significant association between ridesourcing entry and road injuries. Our findings are also aligned with studies that found no significant association between ridesourcing entry and fatalities (e.g.,~\cite{Brazil2016,Huang2019}), while we also account for the ridesourcing induced demand effects that the aforementioned studies do not capture. 

The magnitude of the percentage decrease of road safety externalities associated with ridesourcing use can be considered small compared to the effectiveness of other strategic road safety interventions. Seatbelt laws enforcement may result in up to 9\% reduction in road injuries~\cite{Carpenter2008} while speed limit reduction and traffic calming measures may enable to reach 10-15\% decrease in traffic crashes~\cite{Elvik2001}. Therefore, promoting the use of ridesourcing as vision zero policy might not be as effective as prioritizing infrastructure improvements, speed limit changes, and other educational or policing vision zero initiatives. However, a ridesourcing initiative to reduce DWI offenses and increase road safety could benefit groups most likely to ride those such as younger drivers~\cite{Brazil2016}.

Road crashes and fatalities are significantly associated with population density, as expected. Crashes increase with increased population density, but road fatalities decrease. Population density as fatalities' predictor might serve as a proxy for speed's effect, given that roads in lower density environments tend to be characterized by higher speed limits. The percentage of vehicle owners is positively associated with road crashes and injuries in Travis County. This result is aligned with other literature findings~\cite{Barrios2019}; lower levels of vehicle ownership hint at greater public transit and/or active transportation usage. We note, though, that ridesourcing use could endogenously influence vehicle ownership positively by incentivizing more to drive for transportation network companies. In our analysis, ridesourcing use is positively (but weakly) correlated (corr$=0.12$) with the percentage of zero vehicle ownership (shown in \nameref{S1_Fig}).

The number of DWI offenses would decrease when a greater number of alternative transportation options are available, as expected. This finding also suggests that ridesourcing in Travis County may not only serve as a substitute for taxis and other modes of transportation but also for drunk driving. However, this outcome could also be a result of enforcement, educational, and other efforts, which we are unable to capture here. 

\subsection*{Robustness checks}
The concurrent launch of RideAustin and the exit of Uber and Lyft from the Austin region~\cite{Solomon2017} during the period of May and June of 2016 might result in skewing our modeling outcomes since the association of ridesourcing use with road crashes, DWI offenses, and other road safety indices might be attributed to the departure of the popular ridesourcing services. We conduct a robustness check to examine whether the magnitude and sign of the ridesourcing use parameter remain the same when using only the last six months of the ride data of our RideAustin treatment period. This enables us to exclude data representative of the initial and transitory RideAustin operation period and focus only on the operation between October 2016 to March 2017, assuming that the impacts of the exit of Uber and Lyft will have subsided. \nameref{S1_Table}, \nameref{S2_Table}, and \nameref{S3_Table} showcase modeling results of the spatial lag, spatial error, and SARAR models in the SI. Even though the significance of the relationship between ridesourcing use and road safety outcomes is weaker in this case, we observe that the sign of the parameter remains negative. In accordance with the base RideAustin operating period definition, we observe that ridesourcing use is associated with a reduction in DWI offenses and traffic injuries but does not seem to be related to road crashes and fatalities. 

Note that existing literature findings suggest that once Uber and Lyft services depart from Austin, the crashes and DWI offenses decline~\cite{Barrios2019}. In such a case, some of the observed declines in road crashes might be attributed to reductions in the number of cars on the road. Therefore, the treatment might be the departure of Uber and Lyft, not the arrival of RideAustin. In order to be able to explore such a notion, we would need information about the trips conducted by ridesourcing services over the excluded period of analysis. 

\section*{Conclusion}
We use RideAustin trips to examine the effect of ridesourcing exposure on road safety outcomes, such as road crashes, injuries, fatalities, and DWI offenses. Spatial fixed-effects panel data models are employed to establish that RideAustin use is significantly associated with a decrease in total road crashes, injuries, and DWI offenses in Travis County, Texas. On the contrary, our findings do not demonstrate significant relationships between ridesourcing use and road fatalities. Given the significant costs associated with road safety outcomes and DWI offenses~\cite{DOTHS}, ridesourcing services can be a low-cost option that could assist cities and counties with meeting goals for road injuries and DWI offenses reduction. At the same time, the magnitude of the road safety externalities decrease that is associated with ridesourcing trips is smaller compared to the safety effectiveness that has been documented after the application of other interventions, including seatbelt laws, reduced speeds, and traffic calming design. This outlines the need for determining population segments that ridesourcing-related solutions or policies could be more impactful for improving road safety, like focusing on younger ridesourcing demographics.

Our analysis augments existing work in this field by accounting for spatial distributions of ridesourcing use, road safety outcomes, and other socio-economic characteristics in the given region. Instead of testing associations of the launch of ridesourcing with road injuries and the rest of safety outcomes, we account for spatio-temporal characteristics and capture actual ridesourcing use via real-time trip data analytics in Travis County. The spatial panel data modeling outcomes show that spatial dependence is of significance. Thus, granular longitudinal travel, road safety, and socio-demographic panel data can provide transportation and traffic safety agencies that opportunity to uncover associations and plan for appropriate safety interventions.

Additional research efforts should be put towards addressing this study's limitations, including 1) testing the effect of alternate travel demand exposure methods, such as vehicle miles traveled instead of OD trips, 2) exploring whether effects might not manifest immediately from ridesourcing use, accounting for lags~\cite{Greenwood2017}, 3) examining results robustness by performing similar analysis in additional regions in the US and around the world, since the generalizability of the results can be questioned,  and 4) fitting conditional autoregressive models and commenting on their performance differences compared to spatial autoregressive ones~\cite{Quddus2008}. Future research should also uncover for which populations and subpopulations road safety outcomes can be improved through ridesourcing use by (a) exploring the relationship of ridesourcing and road safety outcomes for different household income and employment percentage panels and (b) identifying critical drivers of where potential public health benefits of ridesourcing utilization can be the greatest. 

\section*{Supporting information}
\paragraph*{S1 Fig.}
\label{S1_Fig}
\begin{figure}[h]
\centering
\includegraphics[width=\textwidth]{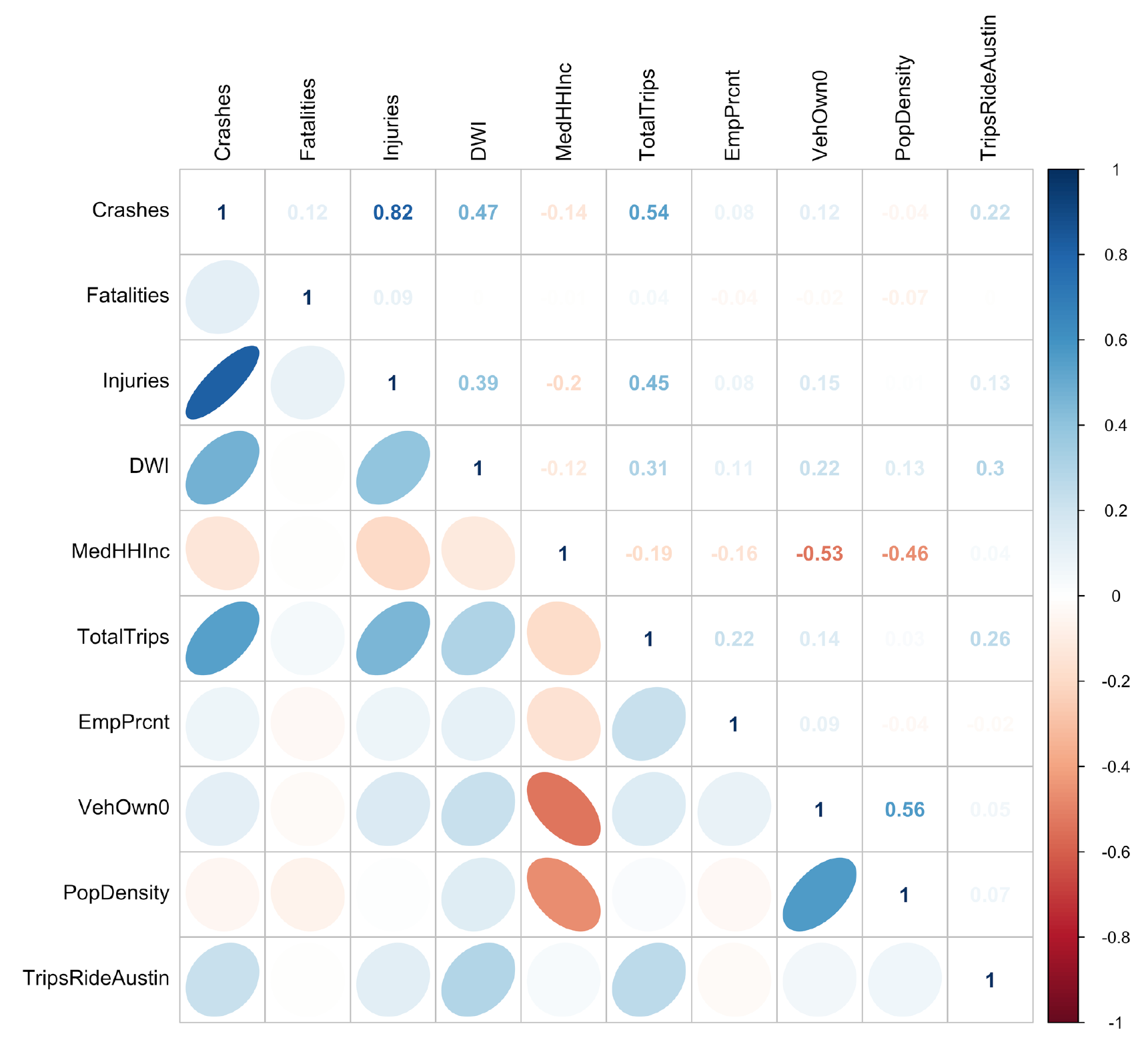}
\end{figure}
{\bf Road crash rate and ridesourcing rates before and after the introduction of the ridesourcing service.} The correlation matrix results of the four safety outcome variables, the ridesourcing use, and control variables are presented to showcase associations between those.

\paragraph*{S1 File.}
\label{S1_File}
{\bf Covariate raw data.} Transportation and socio-demographic data used in the analysis.

\paragraph*{S1 Appendix.}
\label{S1_Appendix}
{\bf Traffic time series analysis.} Due to unavailability of OD traffic data for each census tract in Travis County for the whole period of the analysis (January 2012-June 2014 and June 2016-April 2017), a time series, autoregressive integrated moving average model was fitted. We have monthly OD trips rate output data from the StreetLight Data platform from January 2016 until December 2018~\cite{StreetLight}. The ridesourcing trip rates refer to both origins and destinations for trips that start and end at different census tracts. If both trip's origin and destination fall within the same census tract then it is only counted once. We apply the time series model with a transfer function (based on historical gasoline prices) to populate the historical OD trips data for every spatial unit of analysis. The autoregressive order of the model ARIMA$(p,q,d)$ dictates the number of the lagged values $x_{i,t-1}$ that have an impact on $x_{i,t}$ OD trips. The order of the moving average direction $q$ is the number of the lagged error terms used in the model, and $d$ is the fractional integration parameter used to force stationarity. In our analysis $p=1, q=0, and d=1$. The transfer function is the average monthly \$ per gallon gas price. The chosen model was based on comparisons of R squared adjusted and Akaike's Information Criterion. Its results are used to populate the OD Trips data fields of this analysis.

\paragraph*{S1 Table.}
\label{S1_Table}
{\bf Robustness check using shorter RideAustin operational period dataset (October 2016-March 2017): spatial lag fixed-effects model results.} 
\begin{table}[ht]
\begin{adjustwidth}{-2.25in}{0in}
\centering
\begin{tabular}{|l|ll|ll|ll|ll|}
\hline
                                  & \multicolumn{2}{l|}{\textbf{Log(1+Crashes)}}       & \multicolumn{2}{l|}{\textbf{Log(1+Injuries)}}    & \multicolumn{2}{l|}{\textbf{Log(1+Fatalities)}}   & \multicolumn{2}{l|}{\textbf{Log(1+DWI)}}          \\
                                  & \textbf{$\beta$}                    & \textbf{}    & \textbf{$\beta$}                    & \textbf{}  & \textbf{$\beta$}                    & \textbf{}   & \textbf{$\beta$}                    & \textbf{}   \\ \hline
\thickhline
\textbf{Percent of employment}    & \multicolumn{1}{l|}{-0.042}         &              & \multicolumn{1}{l|}{0.448}          &            & \multicolumn{1}{l|}{-0.096}         & \textbf{}   & \multicolumn{1}{l|}{0.051}          &             \\
                                  & \multicolumn{1}{l|}{[0.179]}        &              & \multicolumn{1}{l|}{[0.242]}        &            & \multicolumn{1}{l|}{[0.057]}        &             & \multicolumn{1}{l|}{[0.171]}        &             \\ \hline
\textbf{Median HH income}         & \multicolumn{1}{l|}{$-2.49 10^{-6}$}  & \textbf{.}   & \multicolumn{1}{l|}{$-2.57 10^{-6}$}  &            & \multicolumn{1}{l|}{$0.33 10^{-6}$}   & \textbf{}   & \multicolumn{1}{l|}{$-1.09 10^{-6}$}  & \textbf{}   \\
                                  & \multicolumn{1}{l|}{[$1.13 10^{-6}$]} &              & \multicolumn{1}{l|}{[$1.54 10^{-6}$]} &            & \multicolumn{1}{l|}{[$0.36 10^{-6}$]} &             & \multicolumn{1}{l|}{[$1.09 10^{-6}$]} &             \\ \hline
\textbf{Percent of zero vehicle ownership} & \multicolumn{1}{l|}{-0.689}         & \textbf{.}   & \multicolumn{1}{l|}{-0.675}         & \textbf{}  & \multicolumn{1}{l|}{0.145}          &             & \multicolumn{1}{l|}{-0.114}         &             \\
                                  & \multicolumn{1}{l|}{[0.342]}        &              & \multicolumn{1}{l|}{[0.461]}        &            & \multicolumn{1}{l|}{[0.101]}        &             & \multicolumn{1}{l|}{[0.328]}        &             \\ \hline
\textbf{Population density}                & \multicolumn{1}{l|}{$2.52 10^{-6}$}   & \textbf{}    & \multicolumn{1}{l|}{$2.54 10^{-6}$}   &            & \multicolumn{1}{l|}{$-1.24 10^{-6}$}  & \textbf{**} & \multicolumn{1}{l|}{$1.64 10^{-6}$}   &             \\
                                  & \multicolumn{1}{l|}{[$1.48 10^{-6}$]} &              & \multicolumn{1}{l|}{[$2.00 10^{-6}$]} &            & \multicolumn{1}{l|}{[$0.47 10^{-6}$]} &             & \multicolumn{1}{l|}{[$1.42 10^{-6}$]} &             \\ \hline
\textbf{OD trips}                          & \multicolumn{1}{l|}{$7.78 10^{-7}$}   &              & \multicolumn{1}{l|}{$1.45 10^{-6}$}   &            & \multicolumn{1}{l|}{$0.16 10^{-6}$}   &             & \multicolumn{1}{l|}{$1.24 10^{-6}$}   & \textbf{}   \\
                                  & \multicolumn{1}{l|}{[$7.18 10^{-7}$]} &              & \multicolumn{1}{l|}{[$0.97 10^{-6}$]} &            & \multicolumn{1}{l|}{[$0.23 10^{-6}$]} &             & \multicolumn{1}{l|}{[$0.69 10^{-6}$]} &             \\ \hline
\textbf{Log(1+trips RideAustin)}           & \multicolumn{1}{l|}{-0.011}         & \textbf{}    & \multicolumn{1}{l|}{-0.028}         & \textbf{*} & \multicolumn{1}{l|}{-0.002}         &             & \multicolumn{1}{l|}{-0.029}         & \textbf{**} \\
                                  & \multicolumn{1}{l|}{[0.008]}        &              & \multicolumn{1}{l|}{[0.014]}        &            & \multicolumn{1}{l|}{[0.003]}        &             & \multicolumn{1}{l|}{[0.008]}        &             \\ \hline
\textbf{$\lambda$}                         & \multicolumn{1}{l|}{0.101}          & \textbf{***} & \multicolumn{1}{l|}{0.005}          & \textbf{*}          & \multicolumn{1}{l|}{0.008}          & \textbf{}   & \multicolumn{1}{l|}{0.042}          & \textbf{*}           \\
                                  & \multicolumn{1}{l|}{[0.018]}        &              & \multicolumn{1}{l|}{[0.019]}        &            & \multicolumn{1}{l|}{[0.019]}        &             & \multicolumn{1}{l|}{[0.019]}        &             \\ \hline
\textbf{LM test (df=1)}                    & \multicolumn{1}{l|}{29.35}          & \textbf{***} & \multicolumn{1}{l|}{8.71}           & \textbf{*} & \multicolumn{1}{l|}{0.19}           & \textbf{}   & \multicolumn{1}{l|}{4.85}           & \textbf{.}   \\ \hline
\end{tabular}
\begin{flushleft}Symbol \textbf{***} corresponds to p$<0.0001$, \textbf{**} to p$<0.001$, \textbf{*} to p$<0.01$, and \textbf{.} to p$<0.05$.
\end{flushleft}
\end{adjustwidth}
\end{table}

\newpage

\paragraph*{S2 Table.}
\label{S2_Table}
{\bf Robustness check using shorter RideAustin operational period dataset (October 2016-March 2017): spatial error fixed-effects model results.} 
\begin{table}[!ht]
\begin{adjustwidth}{-2.25in}{0in}
\centering
\begin{tabular}{|l|ll|ll|ll|ll|}
\hline
                                  & \multicolumn{2}{l|}{\textbf{Log(1+Crashes)}}       & \multicolumn{2}{l|}{\textbf{Log(1+Injuries)}}     & \multicolumn{2}{l|}{\textbf{Log(1+Fatalities)}}   & \multicolumn{2}{l|}{\textbf{Log(1+DWI)}}          \\
                                  & \textbf{$\beta$}                    & \textbf{}    & \textbf{$\beta$}                    & \textbf{}   & \textbf{$\beta$}                    & \textbf{}   & \textbf{$\beta$}                    & \textbf{}   \\ \hline
\thickhline
\textbf{Percent of employment}    & \multicolumn{1}{l|}{0.054}          &              & \multicolumn{1}{l|}{0.453}          &             & \multicolumn{1}{l|}{-0.096}         & \textbf{}   & \multicolumn{1}{l|}{0.053}          &             \\
                                  & \multicolumn{1}{l|}{[0.179]}        &              & \multicolumn{1}{l|}{[0.241]}        &             & \multicolumn{1}{l|}{[0.057]}        &             & \multicolumn{1}{l|}{[0.172]}        &             \\ \hline
\textbf{Median HH income}         & \multicolumn{1}{l|}{$-2.45 10^{-6}$}  & \textbf{.}   & \multicolumn{1}{l|}{$-2.55 10^{-6}$}  &             & \multicolumn{1}{l|}{$0.34 10^{-6}$}   & \textbf{.}  & \multicolumn{1}{l|}{$-1.06 10^{-6}$}  & \textbf{}   \\
                                  & \multicolumn{1}{l|}{[$1.14 10^{-6}$]} &              & \multicolumn{1}{l|}{[$1.53 10^{-6}$]} &             & \multicolumn{1}{l|}{[$0.36 10^{-6}$]} &             & \multicolumn{1}{l|}{[$1.09 10^{-6}$]} &             \\ \hline
\textbf{Percent of zero vehicle ownership} & \multicolumn{1}{l|}{-0.655}         & \textbf{}    & \multicolumn{1}{l|}{-0.675}         & \textbf{}   & \multicolumn{1}{l|}{0.146}          &             & \multicolumn{1}{l|}{-0.106}         &             \\
                                  & \multicolumn{1}{l|}{[0.344]}        &              & \multicolumn{1}{l|}{[0.464]}        &             & \multicolumn{1}{l|}{[0.109]}        &             & \multicolumn{1}{l|}{[0.329]}        &             \\ \hline
\textbf{Population density}                & \multicolumn{1}{l|}{$2.41 10^{-6}$}   & \textbf{}    & \multicolumn{1}{l|}{$2.41 10^{-6}$}   &             & \multicolumn{1}{l|}{$-1.24 10^{-6}$}  & \textbf{**} & \multicolumn{1}{l|}{$1.63 10^{-6}$}   &             \\
                                  & \multicolumn{1}{l|}{[$1.48 10^{-6}$]} &              & \multicolumn{1}{l|}{[$2.00 10^{-6}$]} &             & \multicolumn{1}{l|}{[$0.47 10^{-6}$]} &             & \multicolumn{1}{l|}{[$1.42 10^{-6}$]} &             \\ \hline
\textbf{OD trips}                          & \multicolumn{1}{l|}{$7.83 10^{-7}$}   &              & \multicolumn{1}{l|}{$1.45 10^{-6}$}   &             & \multicolumn{1}{l|}{$0.16 10^{-6}$}   &             & \multicolumn{1}{l|}{$1.24 10^{-6}$}   & \textbf{}   \\
                                  & \multicolumn{1}{l|}{[$7.41 10^{-7}$]} &              & \multicolumn{1}{l|}{[$0.98 10^{-6}$]} &             & \multicolumn{1}{l|}{[$0.23 10^{-6}$]} &             & \multicolumn{1}{l|}{[$0.69 10^{-6}$]} &             \\ \hline
\textbf{Log(1+trips RideAustin)}           & \multicolumn{1}{l|}{-0.011}         & \textbf{}    & \multicolumn{1}{l|}{-0.028}         & \textbf{*}  & \multicolumn{1}{l|}{-0.002}         &             & \multicolumn{1}{l|}{-0.029}         & \textbf{**} \\
                                  & \multicolumn{1}{l|}{[0.009]}        &              & \multicolumn{1}{l|}{[0.012]}        &             & \multicolumn{1}{l|}{[0.002]}        &             & \multicolumn{1}{l|}{[0.008]}        &             \\ \hline
\textbf{$\rho$}                            & \multicolumn{1}{l|}{0.100}          & \textbf{***} & \multicolumn{1}{l|}{0.055}          & \textbf{**} & \multicolumn{1}{l|}{0.008}          & \textbf{}   & \multicolumn{1}{l|}{0.041}          & \textbf{.}           \\
                                  & \multicolumn{1}{l|}{[0.018]}        &              & \multicolumn{1}{l|}{[0.019]}        &             & \multicolumn{1}{l|}{[0.019]}        &             & \multicolumn{1}{l|}{[0.019]}        &             \\ \hline
\textbf{LM test (df=1)}                    & \multicolumn{1}{l|}{28.54}          & \textbf{***} & \multicolumn{1}{l|}{8.52}           & \textbf{.}  & \multicolumn{1}{l|}{0.19}           & \textbf{}   & \multicolumn{1}{l|}{4.85}           & \textbf{.}  \\ \hline
\end{tabular}
\begin{flushleft}Symbol \textbf{***} corresponds to p$<0.0001$, \textbf{**} to p$<0.001$, \textbf{*} to p$<0.01$, and \textbf{.} to p$<0.05$.
\end{flushleft}
\end{adjustwidth}
\end{table}

\paragraph*{S3 Table.}
\label{S3_Table}
{\bf Robustness check using shorter RideAustin operational period dataset (October 2016-March 2017): SARAR model results.}

\begin{table}[!ht]
\begin{adjustwidth}{-2.25in}{0in}
\centering

\begin{tabular}{|l|ll|ll|ll|ll|}
\hline
\textbf{}                                & \multicolumn{2}{l|}{\textbf{Log(1+Crashes)}}       & \multicolumn{2}{l|}{\textbf{Log(1+Injuries)}}      & \multicolumn{2}{l|}{\textbf{Log(1+Fatalities)}}    & \multicolumn{2}{l|}{\textbf{Log(1+DWI)}}           \\
\textbf{}                                & \textbf{$\beta$}                    & \textbf{}    & \textbf{$\beta$}                    & \textbf{}    & \textbf{$\beta$}                    & \textbf{}    & \textbf{$\beta$}                    & \textbf{}    \\ \hline
\textbf{Percent of employment}           & \multicolumn{1}{l|}{-0.019}         &              & \multicolumn{1}{l|}{0.438}          &              & \multicolumn{1}{l|}{-0.092}         & \textbf{}    & \multicolumn{1}{l|}{0.051}          &              \\
\textbf{}                                & \multicolumn{1}{l|}{[0.175]}        &              & \multicolumn{1}{l|}{[0.241]}        &              & \multicolumn{1}{l|}{[0.056]}        &              & \multicolumn{1}{l|}{[0.172]}        &              \\ \hline
\textbf{Median HH income}                & \multicolumn{1}{l|}{$-2.52 10^{-6}$}  & \textbf{.}   & \multicolumn{1}{l|}{$-2.60 10^{-6}$}  &              & \multicolumn{1}{l|}{$0.38 10^{-6}$}   & \textbf{}    & \multicolumn{1}{l|}{$-1.10 10^{-6}$}  & \textbf{}    \\
\textbf{}                                & \multicolumn{1}{l|}{[$1.12 10^{-6}$]} &              & \multicolumn{1}{l|}{[$1.53 10^{-6}$]} &              & \multicolumn{1}{l|}{[$0.36 10^{-6}$]} &              & \multicolumn{1}{l|}{[$1.09 10^{-6}$]} &              \\ \hline
\textbf{Percent of zero vehicle ownership}        & \multicolumn{1}{l|}{-0.743}         & \textbf{.}   & \multicolumn{1}{l|}{-0.703}         & \textbf{}    & \multicolumn{1}{l|}{0.151}          &              & \multicolumn{1}{l|}{-0.118}         &              \\
\textbf{}                                & \multicolumn{1}{l|}{[0.329]}        &              & \multicolumn{1}{l|}{[0.456]}        &              & \multicolumn{1}{l|}{[0.110]}        &              & \multicolumn{1}{l|}{[0.327]}        &              \\ \hline
\textbf{Population density}                       & \multicolumn{1}{l|}{$2.73 10^{-6}$}   & \textbf{.}   & \multicolumn{1}{l|}{$2.76 10^{-6}$}   &              & \multicolumn{1}{l|}{$-1.28 10^{-6}$}  & \textbf{**}  & \multicolumn{1}{l|}{$1.64 10^{-6}$}   &              \\
\textbf{}                                & \multicolumn{1}{l|}{[$1.46 10^{-6}$]} &              & \multicolumn{1}{l|}{[$1.99 10^{-6}$]} &              & \multicolumn{1}{l|}{[$0.46 10^{-6}$]} &              & \multicolumn{1}{l|}{[$1.42 10^{-6}$]} &              \\ \hline
\textbf{OD Trips}                        & \multicolumn{1}{l|}{$0.64 10^{-6}$}   &              & \multicolumn{1}{l|}{$1.42 10^{-6}$}   &              & \multicolumn{1}{l|}{$0.18 10^{-6}$}   &              & \multicolumn{1}{l|}{$1.22 10^{-6}$}   & \textbf{.}   \\
\textbf{}                                & \multicolumn{1}{l|}{[$0.66 10^{-6}$]} &              & \multicolumn{1}{l|}{[$0.94 10^{-6}$]} &              & \multicolumn{1}{l|}{[$0.24 10{-6}$]}  &              & \multicolumn{1}{l|}{[$0.72 10^{-6}$]} &              \\ \hline
\textbf{Log(1+trips RideAustin)}         & \multicolumn{1}{l|}{-0.009}         & \textbf{}    & \multicolumn{1}{l|}{-0.026}         & \textbf{*}   & \multicolumn{1}{l|}{-0.002}         &              & \multicolumn{1}{l|}{-0.028}         & \textbf{**}  \\
\textbf{}                                & \multicolumn{1}{l|}{[0.007]}        &              & \multicolumn{1}{l|}{[0.011]}        &              & \multicolumn{1}{l|}{[0.003]}        &              & \multicolumn{1}{l|}{[0.011]}        &              \\ \hline
\textbf{$\lambda$}                       & \multicolumn{1}{l|}{0.307}          & \textbf{***} & \multicolumn{1}{l|}{0.154}          &              & \multicolumn{1}{l|}{-0.284}         & \textbf{*}   & \multicolumn{1}{l|}{0.059}          &              \\
\textbf{}                                & \multicolumn{1}{l|}{[0.076]}        &              & \multicolumn{1}{l|}{[0.171]}        &              & \multicolumn{1}{l|}{[0.096]}        &              & \multicolumn{1}{l|}{[0.418]}        &              \\ \hline
\textbf{$\rho$}                          & \multicolumn{1}{l|}{-0.236}         & \textbf{.}   & \multicolumn{1}{l|}{-0.105}         &              & \multicolumn{1}{l|}{0.267}          & \textbf{***} & \multicolumn{1}{l|}{-0.017}         &              \\
\textbf{}                                & \multicolumn{1}{l|}{[0.094]}        & \textbf{}    & \multicolumn{1}{l|}{[0.188]}        & \textbf{}    & \multicolumn{1}{l|}{[0.079]}        & \textbf{}    & \multicolumn{1}{l|}{[0.043]}        & \textbf{}    \\ \hline
\textbf{LM: lag (df=1)}                  & \multicolumn{1}{l|}{7.54}           & \textbf{*}   & \multicolumn{1}{l|}{1.04}           & \textbf{}    & \multicolumn{1}{l|}{1.16}           & \textbf{}    & \multicolumn{1}{l|}{3.35}           & \textbf{}    \\ \hline
\textbf{LM: error (df=1)}                & \multicolumn{1}{l|}{6.73}           & \textbf{*}   & \multicolumn{1}{l|}{0.86}           & \textbf{}    & \multicolumn{1}{l|}{1.19}           & \textbf{}    & \multicolumn{1}{l|}{3.08}           & \textbf{}    \\ \hline
\textbf{Hausman test (df=6) chi-squared} & \multicolumn{1}{l|}{139.51}         & \textbf{***} & \multicolumn{1}{l|}{92.48}          & \textbf{***} & \multicolumn{1}{l|}{7.60}           & \textbf{.}            & \multicolumn{1}{l|}{121.85}         & \textbf{***} \\ \hline
\end{tabular}
\begin{flushleft}Symbol \textbf{***} corresponds to p$<0.0001$, \textbf{**} to p$<0.001$, \textbf{*} to p$<0.01$, and \textbf{.} to p$<0.05$.
\end{flushleft}
\end{adjustwidth}
\end{table}


%
%
%


\begin{thebibliography}{10}

\bibitem{Lavieri2018}
Lavieri PS, Dias FF, Juri NR, Kuhr J, Bhat CR.
\newblock A Model of Ridesourcing Demand Generation and Distribution.
\newblock Transportation Research Record: Journal of the Transportation
  Research Board. 2018; p. 036119811875662.
\newblock doi:{10.1177/0361198118756628}.

\bibitem{Hall2018}
Hall JD, Palsson C, Price J.
\newblock Is Uber a substitute or complement for public transit?
\newblock Journal of Urban Economics. 2018;108:36--50.
\newblock doi:{10.1016/j.jue.2018.09.003}.

\bibitem{Ward2019}
Ward JW, Michalek JJ, Azevedo IL, Samaras C, Ferreira P.
\newblock Effects of on-demand ridesourcing on vehicle ownership, fuel
  consumption, vehicle miles traveled, and emissions per capita in U.S. States.
\newblock Transportation Research Part C: Emerging Technologies.
  2019;108:289--301.
\newblock doi:{10.1016/j.trc.2019.07.026}.

\bibitem{Moskatel2019}
Moskatel L, Slusky D.
\newblock Did UberX Reduce Ambulance Volume?
\newblock Health Economics. 2019;28(7):817--829.
\newblock doi:{https://doi.org/10.1002/hec.3888}.

\bibitem{Wolfe2020}
Wolfe MK, McDonald NC.
\newblock Innovative health care mobility services in the US.
\newblock BMC public health. 2020;20:906.
\newblock doi:{10.1186/s12889-020-08803-5}.

\bibitem{Erhardt2019}
Erhardt GD, Roy S, Cooper D, Sana B, Chen M, Castiglione J.
\newblock Do transportation network companies decrease or increase congestion?
\newblock Science Advances. 2019;5.
\newblock doi:{10.1126/sciadv.aau2670}.

\bibitem{Kong2020}
Kong H, Zhang X, Zhao J.
\newblock How does ridesourcing substitute for public transit? A geospatial
  perspective in Chengdu, China.
\newblock Journal of Transport Geography. 2020;86:102769.
\newblock doi:{10.1016/j.jtrangeo.2020.102769}.

\bibitem{Wenzel2019}
Wenzel T, Rames C, Kontou E, Henao A.
\newblock Travel and energy implications of ridesourcing service in Austin,
  Texas.
\newblock Transportation Research Part D: Transport and Environment.
  2019;70:18--34.
\newblock doi:{10.1016/j.trd.2019.03.005}.

\bibitem{Fry2019}
Fry D, Kioumourtzoglou MA, Treat CA, Burke KR, Evans D, Tabb LP, et~al.
\newblock Development and validation of a method to quantify benefits of
  clean-air taxi legislation.
\newblock Journal of Exposure Science \& Environmental Epidemiology.
  2019;doi:{10.1038/s41370-019-0141-6}.

\bibitem{Greenwood2017}
Greenwood BN, Wattal S.
\newblock Show Me the Way to Go Home: An Empirical Investigation of
  Ride-Sharing and Alcohol Related Motor Vehicle Fatalities.
\newblock Management Information Systems Quarterly. 2017;41:163--187.

\bibitem{Goodspeed2019}
Goodspeed R, Xie T, Dillahunt TR, Lustig J.
\newblock An alternative to slow transit, drunk driving, and walking in bad
  weather: An exploratory study of ridesourcing mode choice and demand.
\newblock Journal of Transport Geography. 2019;79:102481.
\newblock doi:{10.1016/j.jtrangeo.2019.102481}.

\bibitem{Kontou2020}
Kontou E, Garikapati V, Hou Y.
\newblock Reducing ridesourcing empty vehicle travel with future travel demand
  prediction.
\newblock Transportation Research Part C. 2020;121:102826.
\newblock doi:{10.1016/j.trc.2020.102826}.

\bibitem{Lavieri2019}
Lavieri PS, Bhat CR.
\newblock Investigating objective and subjective factors influencing the
  adoption , frequency , and characteristics of ride-hailing trips.
\newblock Transportation Research Part C. 2019;105:100--125.
\newblock doi:{10.1016/j.trc.2019.05.037}.

\bibitem{Barrios2019}
Barrios JM, Hochberg YV, Yi LH.
\newblock The Cost of Convenience: Ridesharing and Traffic Fatalities.
  2019;doi:{10.2139/ssrn.3259965}.

\bibitem{Brazil2016}
Brazil N, Kirk DS.
\newblock Uber and Metropolitan Traffic Fatalities in the United States.
\newblock American Journal of Epidemiology. 2016;184:192--198.
\newblock doi:{10.1093/aje/kww062}.

\bibitem{Whitelegg1987}
Whitelegg AJ.
\newblock A Geography of Road Traffic Accidents.
\newblock Transactions of the Institute of British Geographers.
  1987;12:161--176.

\bibitem{Noland2005}
Noland RB, Quddus MA.
\newblock Congestion and safety: A spatial analysis of London.
\newblock Transportation Research Part A: Policy and Practice.
  2005;39:737--754.
\newblock doi:{10.1016/j.tra.2005.02.022}.

\bibitem{Morrison2018}
Morrison CN, Jacoby SF, Dong B, Delgado MK, Wiebe DJ.
\newblock Ridesharing and Motor Vehicle Crashes in 4 US Cities: An Interrupted
  Time-Series Analysis.
\newblock American Journal of Epidemiology. 2018;187:224--232.
\newblock doi:{10.1093/aje/kwx233}.

\bibitem{Dills2018}
Dills AK, Mulholland SE.
\newblock Ride-Sharing, Fatal Crashes, and Crime.
\newblock Southern Economic Journal. 2018;84:965--991.
\newblock doi:{10.1002/soej.12255}.

\bibitem{Huang2019}
Huang JY, Majid F, Daku M.
\newblock Estimating effects of Uber ride-sharing service on road
  traffic-related deaths in South Africa: a quasi-experimental study.
\newblock Journal of Epidemiology and Community Health. 2019; p.
  jech--2018--211006.
\newblock doi:{10.1136/jech-2018-211006}.

\bibitem{Kirk2020}
Kirk DS, Cavalli N, Brazil N.
\newblock The implications of ridehailing for risky driving and road accident
  injuries and fatalities.
\newblock Social Science \& Medicine. 2020;250:112793.
\newblock doi:{10.1016/j.socscimed.2020.112793}.

\bibitem{Dalziel1997}
Dalziel JR, Job RFS.
\newblock Motor vehicle accidents, fatigue and optimism bias in taxi drivers.
\newblock Accident Analysis and Prevention. 1997;29:489--494.
\newblock doi:{10.1016/s0001-4575(97)00028-6}.

\bibitem{Wang2019}
Wang Y, Li L, Prato CG.
\newblock The relation between working conditions, aberrant driving behaviour
  and crash propensity among taxi drivers in China.
\newblock Accident Analysis and Prevention. 2019;126:17--24.
\newblock doi:{10.1016/j.aap.2018.03.028}.

\bibitem{La2013}
La QN, Lee AH, Meuleners LB, Duong DV.
\newblock Prevalence and factors associated with road traffic crash among taxi
  drivers in Hanoi, Vietnam.
\newblock Accident Analysis and Prevention. 2013;50:451--455.
\newblock doi:{10.1016/j.aap.2012.05.022}.

\bibitem{Jones2019}
Jones S, Lidbe A, Hainen A.
\newblock What can open access data from India tell us about road safety and
  sustainable development?
\newblock Journal of Transport Geography. 2019;80:102503.
\newblock doi:{10.1016/j.jtrangeo.2019.102503}.

\bibitem{Huang2018}
Huang Y, Wang X, Patton D.
\newblock Examining spatial relationships between crashes and the built
  environment: A geographically weighted regression approach.
\newblock Journal of Transport Geography. 2018;69:221--233.
\newblock doi:{10.1016/j.jtrangeo.2018.04.027}.

\bibitem{Wang2016}
Wang Y, Chau CK, Ng WY, Leung TM.
\newblock A review on the effects of physical built environment attributes on
  enhancing walking and cycling activity levels within residential
  neighborhoods.
\newblock Cities. 2016;50:1--15.
\newblock doi:{10.1016/j.cities.2015.08.004}.

\bibitem{Wang2007}
Wang X, Kockelman KM.
\newblock Specification and estimation of a spatially and temporally
  autocorrelated seemingly unrelated regression model: Application to crash
  rates in China.
\newblock Transportation. 2007;34:281--300.
\newblock doi:{10.1007/s11116-007-9117-9}.

\bibitem{Xie2014}
Xie K, Wang X, Ozbay K, Yang H.
\newblock Crash frequency modeling for signalized intersections in a
  high-density urban road network.
\newblock Analytic Methods in Accident Research. 2014;2:39--51.
\newblock doi:{10.1016/j.amar.2014.06.001}.

\bibitem{Washington2003}
Washington SP, Karlaftis MG, Mannering F.
\newblock Statistical and econometric methods for transportation data analysis.
\newblock Chapman and Hall/CRC; 2003.

\bibitem{Anselin1995}
Anselin L, Florax RJGM.
\newblock Small Sample Properties of Tests for Spatial Dependence in Regression
  Models: Some Further Results.
\newblock Geographical Analysis. 1995; p. 21--74.
\newblock doi:{10.1007/978-3-642-79877-1\_2}.

\bibitem{Anselin1992}
Anselin L, Hudak S.
\newblock Spatial econometrics in practice. A review of software options.
\newblock Regional Science and Urban Economics. 1992;22:509--536.
\newblock doi:{10.1016/0166-0462(92)90042-Y}.

\bibitem{Baltagi2003}
Baltagi BH, Song SH, Koh W.
\newblock Testing panel data regression models with spatial error correlation.
\newblock Journal of Econometrics. 2003;117:123--150.
\newblock doi:{10.1016/S0304-4076(03)00120-9}.

\bibitem{Debarsy2010}
Debarsy N, Ertur C.
\newblock Testing for spatial autocorrelation in a fixed effects panel data
  model.
\newblock Regional Science and Urban Economics. 2010;40:453--470.
\newblock doi:{10.1016/j.regsciurbeco.2010.06.001}.

\bibitem{Millo2012}
Millo G, Piras G.
\newblock Journal of Statistical Software splm: Spatial Panel Data Models in R.
\newblock Journal of Statistical Software. 2012;47.

\bibitem{Hausman1978}
Hausman JA.
\newblock Specification Tests in Econometrics.
\newblock Econometrica. 1978;46:1251--1271.

\bibitem{RideAustin2017}
RideAustin. Ride-Austin-june6-april13; 2017.
\newblock Available from:
  \url{https://data.world/ride-austin/ride-austin-june-6-april-13}.

\bibitem{Zeitlin2019}
Zeitlin M. How Austin’s failed attempt to regulate Uber and Lyft foreshadowed
  today’s ride-hailing controversies; 2019.
\newblock Available from:
  \url{https://www.vox.com/the-highlight/2019/9/6/20851575/uber-lyft-drivers-austin-regulation-rideshare}.

\bibitem{Hampshire2017}
Hampshire R, Simek C, Fabusuyi T, Di X, Chen X.
\newblock Measuring the Impact of an Unanticipated Disruption of Uber/Lyft in
  Austin, TX.
\newblock Transportation Research Board 97th Annual Meeting Proceedings. 2017;
  p. 1--18.
\newblock doi:{10.2139/ssrn.2977969}.

\bibitem{Klop1999}
Klop J, Khattak AJ.
\newblock Factors Influencing Bicycle Crash Severity on Two-Lane, Undivided
  Roadways in North Carolina.
\newblock Transportation Research Record. 1999;1674:78--85.

\bibitem{Solomon2017}
Solomon D. ONE YEAR AFTER FLEEING AUSTIN, UBER AND LYFT PREPARE A FRESH
  INVASION; 2017.
\newblock Available from:
  \url{https://www.wired.com/2017/05/one-year-fleeing-austin-uber-lyft-prepare-fresh-invasion/}.

\bibitem{Clark2017}
Clark WAV, Lisowski W.
\newblock Decisions to move and decisions to stay: life course events and
  mobility outcomes.
\newblock Housing Studies. 2017;32:547--565.
\newblock doi:{10.1080/02673037.2016.1210100}.

\bibitem{Huang2010}
Huang H, Abdel-Aty MA, Darwiche AL.
\newblock County-level crash risk analysis in Florida: Bayesian spatial
  modeling.
\newblock Transportation Research Record. 2010; p. 27--37.
\newblock doi:{10.3141/2148-04}.

\bibitem{Carpenter2008}
Carpenter CS, Stehr M.
\newblock The effects of mandatory seatbelt laws on seatbelt use, motor vehicle
  fatalities, and crash-related injuries among youths.
\newblock Journal of Health Economics. 2008;27:642--662.
\newblock doi:{10.1016/j.jhealeco.2007.09.010}.

\bibitem{Elvik2001}
Elvik R.
\newblock Area-wide urban traffic calming schemes: A meta-analysis of safety
  effects.
\newblock Accident Analysis and Prevention. 2001;33:327--336.
\newblock doi:{10.1016/S0001-4575(00)00046-4}.

\end{thebibliography}


\begin{thebibliography}{10}


\bibitem{UberSec}
Uber Technologies Inc. Uber Technologies Inc Initial Public Offering; 2019.
\newblock Available from:
  \url{https://www.sec.gov/Archives/edgar/data/1543151/000119312519103850/d647752ds1.htm}.
  
\bibitem{Lavieri2018}
Lavieri PS, Dias FF, Juri NR, Kuhr J, Bhat CR.
\newblock A Model of Ridesourcing Demand Generation and Distribution.
\newblock Transportation Research Record: Journal of the Transportation
  Research Board. 2018; p. 036119811875662.
\newblock doi:{10.1177/0361198118756628}.

\bibitem{Hall2018}
Hall JD, Palsson C, Price J.
\newblock Is Uber a substitute or complement for public transit?
\newblock Journal of Urban Economics. 2018;108:36--50.
\newblock doi:{10.1016/j.jue.2018.09.003}.

\bibitem{Ward2019}
Ward JW, Michalek JJ, Azevedo IL, Samaras C, Ferreira P.
\newblock Effects of on-demand ridesourcing on vehicle ownership, fuel
  consumption, vehicle miles traveled, and emissions per capita in U.S. States.
\newblock Transportation Research Part C: Emerging Technologies.
  2019;108:289--301.
\newblock doi:{10.1016/j.trc.2019.07.026}.

\bibitem{Moskatel2019}
Moskatel L, Slusky D.
\newblock Did UberX Reduce Ambulance Volume?
\newblock Health Economics. 2019;28(7):817--829.
\newblock doi:{https://doi.org/10.1002/hec.3888}.

\bibitem{Wolfe2020}
Wolfe MK, McDonald NC.
\newblock Innovative health care mobility services in the US.
\newblock BMC public health. 2020;20:906.
\newblock doi:{10.1186/s12889-020-08803-5}.

\bibitem{Erhardt2019}
Erhardt GD, Roy S, Cooper D, Sana B, Chen M, Castiglione J.
\newblock Do transportation network companies decrease or increase congestion?
\newblock Science Advances. 2019;5.
\newblock doi:{10.1126/sciadv.aau2670}.

\bibitem{Kong2020}
Kong H, Zhang X, Zhao J.
\newblock How does ridesourcing substitute for public transit? A geospatial perspective in Chengdu, China.
\newblock Journal of Transport Geography. 2020;86:102769.
\newblock doi:{10.1016/j.jtrangeo.2020.102769}.

\bibitem{Wenzel2019}
Wenzel T, Rames C, Kontou E, Henao A.
\newblock Travel and energy implications of ridesourcing service in Austin,
  Texas.
\newblock Transportation Research Part D: Transport and Environment.
  2019;70:18--34.
\newblock doi:{10.1016/j.trd.2019.03.005}.

\bibitem{Fry2019}
Fry D, Kioumourtzoglou MA, Treat CA, Burke KR, Evans D, Tabb LP, et~al.
\newblock Development and validation of a method to quantify benefits of
  clean-air taxi legislation.
\newblock Journal of Exposure Science \& Environmental Epidemiology.
  2019;doi:{10.1038/s41370-019-0141-6}.

\bibitem{Greenwood2017}
Greenwood BN, Wattal S.
\newblock Show Me the Way to Go Home: An Empirical Investigation of
  Ride-Sharing and Alcohol Related Motor Vehicle Fatalities.
\newblock Management Information Systems Quarterly. 2017;41:163--187.

\bibitem{UNDESA}
United Nations Department of Economics and Social Affairs \#Envision2030 Goal 11: Sustainable Cities and
  Communities; 2015.
\newblock Available from:
  \url{https://www.un.org/development/desa/disabilities/envision2030-goal11.html}.

\bibitem{Goodspeed2019}
Goodspeed R, Xie T, Dillahunt TR, Lustig J.
\newblock An alternative to slow transit, drunk driving, and walking in bad weather: An exploratory study of ridesourcing mode choice and demand.
\newblock Journal of Transport Geography. 2019;79:102481.
\newblock doi:{10.1016/j.jtrangeo.2019.102481}.

\bibitem{Kontou2020}
Kontou E, Garikapati V, Hou Y.
\newblock Reducing ridesourcing empty vehicle travel with future travel demand prediction.
\newblock Transportation Research Part C. 2020;121:102826.
\newblock doi:{10.1016/j.trc.2020.102826}.

\bibitem{Lavieri2019}
Lavieri PS, Bhat CR.
\newblock Investigating objective and subjective factors influencing the
  adoption , frequency , and characteristics of ride-hailing trips.
\newblock Transportation Research Part C. 2019;105:100--125.
\newblock doi:{10.1016/j.trc.2019.05.037}.

\bibitem{Barrios2019}
Barrios JM, Hochberg YV, Yi LH.
\newblock The Cost of Convenience: Ridesharing and Traffic Fatalities.
  2019;doi:{10.2139/ssrn.3259965}.

\bibitem{Brazil2016}
Brazil N, Kirk DS.
\newblock Uber and Metropolitan Traffic Fatalities in the United States.
\newblock American Journal of Epidemiology. 2016;184:192--198.
\newblock doi:{10.1093/aje/kww062}.

\bibitem{Whitelegg1987}
Whitelegg AJ.
\newblock A Geography of Road Traffic Accidents.
\newblock Transactions of the Institute of British Geographers.
  1987;12:161--176.
  
\bibitem{Noland2005}
Noland RB, Quddus MA.
\newblock Congestion and safety: A spatial analysis of London.
\newblock Transportation Research Part A: Policy and Practice.
  2005;39:737--754.
\newblock doi:{10.1016/j.tra.2005.02.022}.
  
\bibitem{MartinBuck}
Martin-Buck F.
\newblock Driving Safety: An Empirical Analysis of Ridesharing’ s Impact on Drunk Driving and Alcohol-Related Crime. 2016; p. 1--44.
\newblock doi:{10.1111/j.1756-8765.2012.01221.x}.

\bibitem{Morrison2018}
Morrison CN, Jacoby SF, Dong B, Delgado MK, Wiebe DJ.
\newblock Ridesharing and Motor Vehicle Crashes in 4 US Cities: An Interrupted Time-Series Analysis.
\newblock American Journal of Epidemiology. 2018;187:224--232.
\newblock doi:{10.1093/aje/kwx233}.

\bibitem{Huang2019}
Huang JY, Majid F, Daku M.
\newblock Estimating effects of Uber ride-sharing service on road
  traffic-related deaths in South Africa: a quasi-experimental study.
\newblock Journal of Epidemiology and Community Health. 2019; p.
  jech--2018--211006.
\newblock doi:{10.1136/jech-2018-211006}.

\bibitem{Dills2018}
Dills AK, Mulholland SE.
\newblock Ride-Sharing, Fatal Crashes, and Crime.
\newblock Southern Economic Journal. 2018;84:965--991.
\newblock doi:{10.1002/soej.12255}.

\bibitem{Kirk2020}
Kirk DS, Cavalli N, Brazil N.
\newblock The implications of ridehailing for risky driving and road accident injuries and fatalities.
\newblock Social Science \& Medicine. 2020;250:112793.
\newblock doi:{10.1016/j.socscimed.2020.112793}.

\bibitem{Dalziel1997}
Dalziel JR, Job RFS.
\newblock Motor vehicle accidents, fatigue and optimism bias in taxi drivers.
\newblock Accident Analysis and Prevention. 1997;29:489--494.
\newblock doi:{10.1016/s0001-4575(97)00028-6}.

\bibitem{Wang2019}
Wang Y, Li L, Prato CG.
\newblock The relation between working conditions, aberrant driving behaviour and crash propensity among taxi drivers in China.
\newblock Accident Analysis and Prevention. 2019;126:17--24.
\newblock doi:{10.1016/j.aap.2018.03.028}.

\bibitem{La2013}
La QN, Lee AH, Meuleners LB, Duong DV.
\newblock Prevalence and factors associated with road traffic crash among taxi drivers in Hanoi, Vietnam.
\newblock Accident Analysis and Prevention. 2013;50:451--455.
\newblock doi:{10.1016/j.aap.2012.05.022}.

\bibitem{Gonzalez2019}
González SR, Loukaitou-Sideris A, Chapple K.
\newblock Transit neighborhoods, commercial gentrification, and traffic
crashes: Exploring the linkages in Los Angeles and the Bay Area.
\newblock Journal of Transport Geography. 2019;77:79--89.
\newblock doi:{10.1016/j.jtrangeo.2019.04.010}.

\bibitem{Jones2019}
Jones S, Lidbe A, Hainen A.
\newblock What can open access data from India tell us about road safety and sustainable development?
\newblock Journal of Transport Geography. 2019;80:102503.
\newblock doi:{10.1016/j.jtrangeo.2019.102503}.

\bibitem{Huang2018}
Huang Y, Wang X, Patton D.
\newblock Examining spatial relationships between crashes and the built
  environment: A geographically weighted regression approach.
\newblock Journal of Transport Geography. 2018;69:221--233.
\newblock doi:{10.1016/j.jtrangeo.2018.04.027}.

\bibitem{Wang2016}
Wang Y, Chau CK, Ng WY, Leung TM.
\newblock A review on the effects of physical built environment attributes on enhancing walking and cycling activity levels within residential neighborhoods.
\newblock Cities. 2016;50:1--15.
\newblock doi:{10.1016/j.cities.2015.08.004}.

\bibitem{Wang2007}
Wang X, Kockelman KM.
\newblock Specification and estimation of a spatially and temporally
  autocorrelated seemingly unrelated regression model: Application to crash rates in China.
\newblock Transportation. 2007;34:281--300.
\newblock doi:{10.1007/s11116-007-9117-9}.

\bibitem{Xie2014}
Xie K, Wang X, Ozbay K, Yang H.
\newblock Crash frequency modeling for signalized intersections in a
  high-density urban road network.
\newblock Analytic Methods in Accident Research. 2014;2:39--51.
\newblock doi:{10.1016/j.amar.2014.06.001}.

\bibitem{AbdelAty2007}
Abdel-Aty M, Wang X.
\newblock Crash Estimation at Signalized Intersections Along Corridors:
  Analyzing Spatial Effect and Identifying Significant Factors.
\newblock Transportation Research Record: Journal of the Transportation
  Research Board. 2007;1953:98--111.
\newblock doi:{10.3141/1953-12}.

\bibitem{Washington2003}
Washington SP, Karlaftis MG, Mannering F.
\newblock Statistical and econometric methods for transportation data analysis.
\newblock Chapman and Hall/CRC; 2003.

\bibitem{Anselin1995}
Anselin L, Florax RJGM.
\newblock Small Sample Properties of Tests for Spatial Dependence in Regression Models: Some Further Results.
\newblock Geographical Analysis. 1995; p. 21--74.
\newblock doi:{10.1007/978-3-642-79877-1\_2}.

\bibitem{Anselin1992}
Anselin L, Hudak S.
\newblock Spatial econometrics in practice. A review of software options.
\newblock Regional Science and Urban Economics. 1992;22:509--536.
\newblock doi:{10.1016/0166-0462(92)90042-Y}.

\bibitem{Baltagi2003}
Baltagi BH, Song SH, Koh W.
\newblock Testing panel data regression models with spatial error correlation.
\newblock Journal of Econometrics. 2003;117:123--150.
\newblock doi:{10.1016/S0304-4076(03)00120-9}.

\bibitem{Debarsy2010}
Debarsy N, Ertur C.
\newblock Testing for spatial autocorrelation in a fixed-effects panel data model.
\newblock Regional Science and Urban Economics. 2010;40:453--470.
\newblock doi:{10.1016/j.regsciurbeco.2010.06.001}.

\bibitem{Millo2012}
Millo G, Piras G.
\newblock Journal of Statistical Software splm: Spatial Panel Data Models in R.
\newblock Journal of Statistical Software. 2012;47.

\bibitem{Hausman1978}
Hausman JA.
\newblock Specification Tests in Econometrics.
\newblock Econometrica. 1978;46:1251--1271.

\bibitem{TDOT}
Texas Department of Transportation. Texas Department of Transportation Crash Records
  Information System; 2019.
\newblock Available from:
  \url{https://www.txdot.gov/government/enforcement/crash-statistics.html}.

\bibitem{AustinPD}
Austin Police Department. Crime Reports; 2019.
\newblock Available from:
  \url{https://data.austintexas.gov/Public-Safety/Crime-Reports/fdj4-gpfu/data}.

\bibitem{RideAustin2017}
RideAustin. Ride-Austin-june6-april13; 2017.
\newblock Available from:
  \url{https://data.world/ride-austin/ride-austin-june-6-april-13}.

\bibitem{Zeitlin2019}
Zeitlin M. How Austin’s failed attempt to regulate Uber and Lyft foreshadowed today’s ride-hailing controversies; 2019.
\newblock Available from:
  \url{https://www.vox.com/the-highlight/2019/9/6/20851575/uber-lyft-drivers-austin-regulation-rideshare}.
  
\bibitem{Hampshire2017}
Hampshire R, Simek C, Fabusuyi T, Di X, Chen X.
\newblock Measuring the Impact of an Unanticipated Disruption of Uber/Lyft in Austin, TX.
\newblock Transportation Research Board 97th Annual Meeting Proceedings. 2017; p. 1--18.
\newblock doi:{10.2139/ssrn.2977969}.
  
\bibitem{StreetLight}
StreetLight Data. Essentials for Everyday Traffic Analysis; 2019.
\newblock Available from:
  \url{https://www.streetlightdata.com/transportation-metrics/}.  

\bibitem{TDOTRoad}
Texas Department of Transportation. Roadway Inventory Annual Data; 2018.
\newblock Available from:
  \url{https://www.txdot.gov/inside-txdot/division/transportation-planning/roadway-inventory.html}.

\bibitem{ACS}
United States Census Bureau. American Community Survey; 2018.
\newblock Available from: \url{https://www.census.gov/programs-surveys/acs/}.

\bibitem{Klop1999}
Klop J, Khattak AJ.
\newblock Factors Influencing Bicycle Crash Severity on Two-Lane, Undivided Roadways in North Carolina.
\newblock Transportation Research Record. 1999;1674:78--85.

\bibitem{FHWA}
Federal Highway Administration. KABCO Injury Classification Scale and Definitions; 2018.
\newblock Available from:
  \url{https://safety.fhwa.dot.gov/hsip/spm/conversion\_tbl/pdfs/kabco\_ctable\_by\_state.pdf}.

\bibitem{Solomon2017}
Solomon D. One Year after Fleeing Austin, Uber and Lyft Prepare a Fresh
Invasion; 2017.
\newblock Available from:
  \url{https://www.wired.com/2017/05/one-year-fleeing-austin-uber-lyft-prepare-fresh-invasion/}.

\bibitem{Clark2017}
Clark WAV, Lisowski W.
\newblock Decisions to move and decisions to stay: life course events and
  mobility outcomes.
\newblock Housing Studies. 2017;32:547--565.
\newblock doi:{10.1080/02673037.2016.1210100}.

\bibitem{Huang2010}
Huang H, Abdel-Aty MA, Darwiche AL.
\newblock County-level crash risk analysis in Florida: Bayesian spatial
  modeling.
\newblock Transportation Research Record. 2010; p. 27--37.
\newblock doi:{10.3141/2148-04}.

\bibitem{Carpenter2008}
Carpenter CS, Stehr M.
\newblock The effects of mandatory seatbelt laws on seatbelt use, motor vehicle fatalities, and crash-related injuries among youths.
\newblock Journal of Health Economics. 2008;27:642--662.
\newblock doi:{10.1016/j.jhealeco.2007.09.010}.

\bibitem{Elvik2001}
Elvik R.
\newblock Area-wide urban traffic calming schemes: A meta-analysis of safety effects.
\newblock Accident Analysis and Prevention. 2001;33:327--336.
\newblock doi:{10.1016/S0001-4575(00)00046-4}.

\bibitem{DOTHS}
National Highway Traffic Safety Administration, US Department of Transportation.
\newblock TRAFFIC SAFETY FACTS 2015 Motor Vehicle Crashes: Overview. 2017; p. 1--9.

\bibitem{Quddus2008}
Quddus MA.
\newblock Modelling area-wide count outcomes with spatial correlation and heterogeneity: An analysis of London crash data. 
\newblock Accident Analysis and Prevention. 2008;40:1486--1497.
\newblock doi:{10.1016/j.aap.2008.03.009}.

\end{thebibliography}
\section*{Acknowledgments}
We thank StreetLight Data for providing Kontou with academic access to their travel demand modeling platform. We acknowledge the feedback that we received for an earlier iteration of this paper from the statistical and data science consulting service coordinated by Prof. James Stephen Marron and students of the Department of Statistics and Operations Research at the University of North Carolina at Chapel Hill. Kontou acknowledges an honorarium she received from Pervasive Technology Institute of Indiana University, which enabled her to use cloud computing resources for this analysis.

\end{document}